\documentclass[%
preprint,
 amsmath,amssymb,
 aps,
 longbibliography
]{revtex4-2}

\usepackage{graphicx}
\usepackage{dcolumn}
\usepackage{bm}
\usepackage{physics}
\usepackage{tikz}
\usepackage{bigints}
\usepackage{pgfplots}
\pgfplotsset{compat=1.18}
\usepackage{xcolor}
\usepackage{subcaption}
\usepackage{scalerel,amssymb}

\begin{document}
\definecolor{no_dpdx_sgs_cor_color}{HTML}{57CB9F}
\definecolor{no_RS_div_color}{HTML}{DE413E}
\definecolor{no_delta_v_w_color}{HTML}{66439D}
\definecolor{no_diss}{HTML}{fcaf58}

\newcommand{\Bp}{^{B\prime}}

\newcommand{\errnodpdxsgscor}[1][]{\begin{tikzpicture}[#1]
    \draw[no_dpdx_sgs_cor_color, line width=0.5mm] (0,0) -- (1,0);
    \end{tikzpicture}}

\newcommand{\errnoRSdiv}[1][]{\begin{tikzpicture}[#1]
    \draw[no_RS_div_color, line width=0.5mm] (0,0) -- (1,0);
    \end{tikzpicture}}

\newcommand{\errnodeltavw}[1][]{\begin{tikzpicture}[#1]
    \draw[no_delta_v_w_color, line width=0.5mm] (0,0) -- (1,0);
    \end{tikzpicture}}

\newcommand{\errfullLES}[1][]{\begin{tikzpicture}[#1]
    \draw[dotted, black, line width=0.5mm] (0,0) -- (0.9,0);
    \end{tikzpicture}}

\newcommand{\errnodiss}[1][]{\begin{tikzpicture}[#1]
    \draw[no_diss, line width=0.5mm] (0,0) -- (0.9,0);
    \end{tikzpicture}}

\title{Momentum deficit and wake-added turbulence kinetic energy budgets in the stratified atmospheric boundary layer}

\author{Kerry S. Klemmer}
\affiliation{Department of Civil and Environmental Engineering, Massachusetts Institute of Technology, Cambridge, MA 02139, USA}%

\author{Michael F. Howland}%
 \email{mhowland@mit.edu}
\affiliation{Department of Civil and Environmental Engineering, Massachusetts Institute of Technology, Cambridge, MA 02139, USA}%

\date{\today}

\begin{abstract}
To achieve decarbonization targets, wind turbines are growing in hub height, rotor diameter, and are being deployed in new locations with diverse atmospheric conditions not previously seen, such as offshore. Physics-based analytical wake models commonly used for design and control of wind farms simplify atmospheric boundary layer (ABL) and wake physics to achieve computational efficiency. This is done primarily through a simplified model form that neglects certain flow processes and through parameterization of ABL and wake turbulence through a wake spreading rate. In this study, we analyze the physical mechanisms that govern momentum and turbulence within a wind turbine wake in the stratified ABL. We use large eddy simulation and analysis of the streamwise momentum deficit and wake-added turbulence kinetic energy (TKE) budgets to study wind turbine wakes under neutral and stable conditions. To parse the wake from the turbulent, incident ABL flow, we decompose the flow into the base ABL flow and the deficit flow produced by the presence of a turbine. We analyze the decomposed flow field budgets to study the effects of changing stability on the streamwise momentum deficit and wake-added TKE. The results demonstrate that stability changes the importance of physical mechanisms for both quantities primarily through the nonlinear interactions of the base and deficit flows, with the stable case most affected by higher shear in the base flow and the neutral case by higher base flow TKE. Buoyancy forcing terms in the momentum deficit and wake-added TKE budgets are relatively less important compared to the aforementioned effects. While total TKE is higher in wakes in neutral ABL flows, the wake-added TKE is higher downwind of turbines in stable ABL conditions. The dependence of wake-added TKE on ABL stability is not represented in existing empirical models widely used for mean wake flow modeling.

\end{abstract}

\maketitle


\section{\label{sec:intro} Introduction}

Wind turbine wakes are regions of momentum deficit and increased turbulence that arise due to the energy that turbines extract from the incoming wind in the atmospheric boundary layer (ABL) and enhanced mixing. These wakes often result in power losses for collections of wind turbines collocated within wind farms, which are affected by many factors, including farm layout, wind direction, wind speed, and turbulence content in the wake. Wake interactions also depend on the stability of the ABL. The character of the ABL is particularly affected by atmospheric stability, which arises from buoyancy effects~\cite{nieuwstadt_turbulent_1984,stull_introduction_2012,moeng_comparison_1994}. Stability has two primary impacts on the wake, affecting the incident wind conditions---altering the speed and direction shear---to the turbine as well as the buoyancy forcing in the wake. The former represents the indirect effects of stratification on the wake, while the latter is the direct effect. The resulting wake dynamics in the stratified ABL will depend on the balance between these direct and indirect mechanisms. These changes lead to differences in the transport of heat, momentum, and energy in the wake~\cite{cortina_distribution_2016,wu_new_2023}. Stratification in wind turbine wakes has been studied for single and multi-turbine configurations, with and without Coriolis forcing~\cite{wu_new_2023, abkar_influence_2015, cortina_distribution_2016, xie_numerical_2017, ishihara_new_2018,ali_turbulence_2019}. These studies have investigated myriad aspects of the differences between wind turbine wakes in neutral, stable, and unstable stratification, including surface heat transport~\cite{wu_new_2023}, wake spreading~\cite{abkar_influence_2015,wu_new_2023}, and transport of mean kinetic energy~\cite{cortina_distribution_2016} and turbulence kinetic energy (TKE)~\cite{abkar_influence_2015,ali_turbulence_2019}. These studies show that stratification has a substantial effect on wind turbine wakes. Yet despite this, engineering models that are used in practice are typically derived for neutral or even uniform inflow conditions. As turbines grow in rotor diameter, the limitations of stratification-agnostic wake models are exacerbated due to the expected changes in direction and speed shear in the ABL inflow.

In studying the effect of atmospheric stratification on wind turbine wakes, it is important to differentiate between the direct and indirect effects of stratification on the ABL. The direct effects are felt through buoyancy forcing, which acts to suppress turbulence in stable conditions and produce turbulence in unstable conditions. The indirect effects are those that alter the boundary layer structure as a consequence of the direct buoyancy forcing because the boundary layer inflow to the turbine affects its wake. These include changing the degree of direction shear, speed shear, and turbulence, as well as the development of features such as low-level jets (LLJs)~\cite{banta_stableboundarylayer_2008,doosttalab_interaction_2020}. Current practice primarily employs models that are based on one-dimensional momentum theory~\cite{bastankhah_new_2014,niayifar_analytical_2016}, originally derived via streamtube analysis~\cite{rankine_mechanical_1865,froude_elementary_1878,froude_part_1889}, to predict the momentum or velocity deficit in the wake. These are typically analytical models that assume a particular wake shape, such as a self-similar Gaussian profile~\cite{bastankhah_new_2014}, that expands linearly due to turbulent mixing~\cite{stevens_flow_2017}. Turbulence is only incorporated insofar as it acts to replenish the wake through mixing. While this can work well for uniform or neutral inflow~\cite{bastankhah_new_2014}, the assumptions present in these models do not consider the direct or indirect forcing from stratification. When stratification is considered, it has been done through modifications that typically do not drastically alter the model form, and often only partially consider the indirect effects of stratification, such as shear or wake skewing~\cite{abkar_influence_2016,he_novel_2021}. 

As stated above, turbulence is often incorporated in these analytical engineering wake models through a linear wake spreading rate~\cite{stevens_flow_2017}. The wake spreading rate is parametrized, often, as a linear function of the turbulence intensity, which is a combination of the wake-added turbulence intensity and the ambient turbulence intensity~\cite{niayifar_analytical_2016}. These models for the turbulence added to the wake by the turbine are often highly empirical, with one of the most widely used being the model from Crespo and Hern\'andez~\cite{crespo_turbulence_1996}. The empiricism in these models neglects the impact of ABL physics on the turbulence, ignoring stratification and Coriolis forcing. If these additional ABL physics are incorporated, it tends to be through corrections to existing models without interrogating the original model form. For example, Ishihara and Qian proposed a new model for the wake-added turbulence intensity using a self-similar dual-Gaussian profile and fitting the parameters based on data from both neutral and unstable boundary layers~\cite{ishihara_new_2018}. While they found improved results in comparison to other wake-added turbulence models, these results are likely dependent upon the ambient turbulence intensity. In general, models that rely on ambient streamwise turbulence intensity~\cite{niayifar_analytical_2016} often take this quantity to be spatially and temporally constant~\cite{bensason_evaluation_2021, hoek_predicting_2020}.  A recent work from Klemmer et al.~\cite{klemmer_evaluation_2024} found that averaging turbulence intensity over a year---as opposed to 10-minute intervals as is done with wind direction and wind speed---can lead to errors in farm power of 3.5\% and errors in farm efficiency of 5.0\%. These errors can lead to losses for wind farm and grid operators, which further motivates the need for higher fidelity turbulence models that are derived based on the flow physics, while still maintaining computational efficiency.

While analytical wake models tend to employ the aforementioned empirical turbulence models, there are wake models that use an alternative model form to better capture wake physics. These include dynamic wake meandering (DWM) models~\cite{larsen_dynamic_2007} and variants of the curled wake model~\cite{martinez-tossas_aerodynamics_2019}. Both the DWM model and the curled wake model solve a simplified form of the Navier-Stokes equations to calculate the velocity deficit. The DWM model uses the steady state, axisymmetric thin shear layer approximation of the Navier-Stokes equations and models temporal wake meandering through the interaction of the steady state wake deficit with large-scale turbulent structures~\cite{larsen_dynamic_2007}. An eddy viscosity closure is most commonly used in the DWM model~\cite{keck_implementation_2012,keck_atmospheric_2014}. In Keck et al.~\cite{keck_atmospheric_2014}, the authors address the impact of non-neutral atmospheric stability by tuning the length and velocity scales in the ambient turbulence generated by the Mann turbulence model~\cite{mann_spatial_1994,pena_lengthscale_2010}. The curled wake model solves for the mean wake deficit using a linearized form of the Reynolds-averaged Navier-Stokes (RANS) equations~\cite{martinez-tossas_aerodynamics_2019, martinez-tossas_curled_2021}. As with the DWM model, eddy viscosity turbulence closure is used both with a mixing length formulation~\cite{martinez-tossas_curled_2021} and with a model fit from experimental and simulation data~\cite{scott_evolution_2023}. While an improvement over traditional analytical wake models, the curled wake model and the DWM model both rely on assumptions about the flow physics to achieve their simplified form. In doing so, these models neglect physical processes that may become relevant in non-neutral stratification conditions. 

In the present work, we study wind turbine wakes in stratified ABL conditions through the streamwise momentum deficit and wake-added TKE. In studying the momentum and turbulence forcings that are critical in different ABL stratifications, the goals of this work are twofold: first, we directly analyze two commonly modeled quantities in the wind energy community to better understand the relevant physical mechanisms in the wake and how these change with stratification; second, we can use these results to inform the model form of the wake momentum deficit and wake-added turbulence for variable atmospheric stability. With this analysis, we aim to understand the relevant physical mechanisms in the wake with changing stratification as a precursor to the development of more robust wake models by targeting the wake momentum deficit and the wake-added turbulence directly. 

For this work, we isolate the wake physics by utilizing an approach from Mart\'inez-Tossas et al.~\cite{martinez-tossas_aerodynamics_2019}, which decomposes the flow field such that the base flow without the turbine is removed. In doing so, it is then possible to isolate the wake dynamics and the direct and indirect ways in which stratification alters the physics in the wake. This analysis will identify which forcing terms are critical to the wake momentum deficit and wake-added turbulence to guide us in developing models that are robust to a range of stability conditions. To our knowledge, this wake deficit momentum and wake-added turbulence budget analysis is a novel contribution to the literature that provides information on wake momentum and wake turbulence forcings for different ABL stabilities that are directly relevant to the modeling community. 

The rest of this work is presented as follows: the large eddy simulation (LES) framework and numerical methods are presented in Section~\ref{subsec:methods LES} and the double decomposition and turbulence budgets used throughout the analysis are presented in Section~\ref{subsec:methods double decomp}. The analysis tools, namely the \textit{a priori} model-style budget analysis and the control volume budget analysis are presented in Sections~\ref{subsec:a priori} and~\ref{subsec:control volume}, respectively. The details of the ABL test cases for the neutral and stable boundary layers are given in Section~\ref{subsec:methods abl simulations}. In Section~\ref{subsec:results momentum}, we present the wake deficit streamwise momentum analysis. In Section~\ref{subsec:results tke}, the TKE budget analysis is shown. The results are discussed in Section~\ref{subsec:discussion} followed by a summary and conclusions provided in Section~\ref{sec:conclusions}.

\section{\label{sec:methods} Method}
\subsection{\label{subsec:methods LES} Large eddy simulation}
Large eddy simulations are run for a conventionally neutral boundary layer (CNBL) and a stable boundary layer (SBL). The data are generated using Pad\'eOps (https://github.com/Howland-Lab/PadeOps)~\cite{ghate_subfilterscale_2017,howland_influence_2020}, which is an open-source, pseudo-spectral computational fluid dynamics solver. Fourier collocation is used in the horizontal directions and a sixth-order staggered compact finite difference scheme is used in the vertical direction~\cite{nagarajan_robust_2003}. For the temporal integration, a fourth-order strong stability-preserving variant of a Runge-Kutta scheme is used~\cite{gottlieb_strong_2001}. The filtered incompressible momentum equation with the Boussinesq approximation for buoyancy is given by
\begin{eqnarray}\label{eqn: LES momentum}
    \pdv{u_i}{t} + u_j \pdv{u_i}{x_j} = - \pdv{p}{x_i} - \pdv{\tau_{ij}}{x_j} + f_i + \frac{\delta_{i3}}{Fr^2} \left(\theta - \theta_0 \right) - \frac{2}{Ro} \epsilon_{ijk}\Omega_j \left (u_k -G_k\right),
\end{eqnarray}
where $u_i$ is the velocity in the $x_i$ direction, $t$ is time, $p$ is the non-dimensional pressure, $\tau_{ij}$ is the subgrid-scale (SGS) stress tensor, $G_k$ is the geostrophic wind velocity vector, and $f_i$ is the turbine model forcing. The non-dimensional potential temperature is given by $\theta$, with $\theta_0$ as the reference non-dimensional potential temperature. The Froude number is given by $Fr = U/\sqrt{gL}$, where $U$ is the geostrophic wind speed, $g$ is gravitational acceleration, and $L$ is a dimensional length scale. The Rossby number is given by $Ro = U/(\omega L)$, where $\omega$ is the Coriolis frequency. 

An equation for the filtered non-dimensional potential temperature is also solved:
\begin{eqnarray}
    \pdv{\theta}{t} + u_j \pdv{\theta}{x_j} = -\pdv{q_j^{SGS}}{x_j},
\end{eqnarray}
where $q_j^\text{SGS}$ is the SGS heat flux. For both the SGS stress tensor and the SGS heat flux, the sigma subfilter-scale model is used~\cite{nicoud_using_2011} with a turbulent Prandtl number of 0.4 for the scalar diffusivity.

For all simulations, a fringe region is used in both the streamwise and lateral directions to force the inflow to the desired profile~\cite{nordstrom_fringe_1999}. The inflow is specified via the concurrent precursor method~\cite{stevens_concurrent_2014}, in which two simulations are run concurrently: a primary simulation with the turbine and a precursor simulation of the same domain as the primary but without the turbine. By employing this method, we are able to consider a single turbine in a finite domain.

To study the wake flow physics in different regimes of stratification, we consider two different ABL flows: a conventionally neutral boundary layer and a stable boundary layer, both with a single turbine. The turbine is modeled with the actuator disk model~\cite{calaf_large_2010} and has a diameter of $D=126$ m, a hub height of 90 m, and a thrust coefficient of $C_T=0.75$. Both cases are driven by a geostrophic wind speed of 12 m/s and have thermal and momentum roughness lengths of 10 cm. The streamwise and lateral grid spacing are both 12.5 m, with a streamwise domain length $L_x$ of 4800 m and a lateral domain length $L_y$ of 2400 m. The vertical extent of the neutral case is 2400 m, while the stable case has a vertical extent of 1600 m. Both cases have vertical grid spacing of $\Delta z = 6.25$. The Rossby and Froude numbers based on the rotor diameter and geostrophic wind speed (see Eq.~\ref{eqn: LES momentum}) are 1306 and 0.3414 for both cases, respectively, and the Coriolis frequency is $1.03\times10^{-4}$ s$^{-1}$, which corresponds to a latitude of 45$^{\circ}$ N.

The two ABL cases primarily differ in the surface boundary conditions and initialization of the potential temperature profiles. A constant surface heat flux $\overline{w'\theta'}_w=0$ K m s$^{-1}$ is prescribed in the neutral case. In the stable case, the surface boundary condition is prescribed as a cooling rate of $\frac{\partial \theta}{\partial t} = -0.25$ K hr$^{-1}$, which is found to be more accurate for stable boundary layers than prescription of the surface heat flux~\cite{basu_inconvenient_2008}. In both cases, the initial potential temperature profile is initially set to 301 K up to 700 m for the neutral case and 50 m for the stable case. Above this height, an inversion with a strength of 0.01 K m$^{-1}$ is prescribed for both cases. Further simulation details for both boundary layers are given in Table~\ref{tab:simulation details}.

\begin{table}
\caption{\label{tab:simulation details}
Simulation parameters for stratified the ABL test cases. $L = -\frac{u_*^3 \theta_0}{\kappa g \overline{w'\theta'}_w}$ is the Obukhov length; $u_*$ is the friction velocity; $N_x$, $N_y$, and $N_z$ are the number of grid points in the streamwise, lateral, and vertical directions; and $\Delta x$, $\Delta y$, and $\Delta z$ are the grid spacings in the streamwise, lateral, and vertical directions.}
\begin{ruledtabular}
\begin{tabular}{c|ccccccc}
Case &
$L$ (m) &
$u_*$ (m s$^{-1}$) &
$N_x \times N_y \times N_z$ &
$\Delta x \times \Delta y \times \Delta z$ (m$^3$)\\
\colrule
 CNBL & $\infty$ & 0.52 & $384 \times 192 \times 384$  &  $12.5 \times 12.5  \times 6.25$ \\
 SBL & 118 & 0.36 & $384 \times 192 \times 256$  &  $12.5 \times 12.5 \times  6.25$ \\
\end{tabular}
\end{ruledtabular}
\end{table}

For each case, a spin-up is run with no turbine. Following this, the domain is rotated using a wind angle controlled~\cite{sescu_control_2014}, such that the wind direction at hub height is zero, and the turbine is added. The wind angle controlled is turned off after the rotation during the simulations which collect statistics. The averaging is then performed after roughly 5 flow-through times. The time at which averaging is begun and the duration of averaging is different for both flows due to the differences in the temporal behavior as quantitatively described below. For the CNBL, we exploit the quasi-steady nature of this flow in order to achieve converged statistics. The spin-up is run for 34 hours (roughly two inertial periods, which allows for only minor changes in wind direction after the spin-up during time-averaging), after which the domain is rotated and the turbine is added. After roughly 5 flow-through times, time-averaging is performed over a 15.5-hour period. For the SBL, the flow never truly becomes quasi-steady because of the time-varying surface boundary condition, so we average over a shorter period and follow the best practices in the literature for when to begin averaging~\cite{wu_new_2023}. After 10 hours the domain is rotated and the turbine is added. Time-averaging is performed over a 10-hour period from hour 11 to hour 21.

\subsection{\label{subsec:methods double decomp} Double flow decomposition}
Throughout this work, we decompose the flow field in two ways, following the example of Mart\'{i}nez-Tossas et al.~\cite{martinez-tossas_aerodynamics_2019,martinez-tossas_curled_2021}. First, each flow field variable is decomposed into a base flow component and a wake deficit component. The base flow is the background ABL that would exist in the absence of the wind turbine, and can also be considered as the freestream inflow to the turbine of interest. The base flow variable is the flow field variable taken from the same flow without a turbine (from the precursor simulation), and the wake deficit component comes from the difference between the turbine flow field and the base flow field. For the instantaneous velocity $u_i$, this is given by
\begin{eqnarray}\label{eqn:flow decomp 1}
    u_i = u_i^B + \Delta u_i,
\end{eqnarray}
where $u_i^B$ is the instantaneous base flow velocity and $\Delta u_i$ is the instantaneous wake deficit velocity. The second decomposition is a Reynolds decomposition, in which each variable is decomposed into a mean quantity and a fluctuating quantity (with zero mean), such that for the instantaneous wake velocity deficit, the Reynolds decomposition is given by
\begin{eqnarray}\label{eqn:flow decomp 2}
    \Delta u_i = \overline{\Delta u_i} + \Delta u_i',
\end{eqnarray}
where $\overline{\Delta u_i}$ is the temporal mean wake velocity deficit and $\Delta u_i'$ is the fluctuating wake velocity deficit. 

\subsubsection{\label{subsubsec: streamwise momentum} Streamwise momentum deficit}
It follows that we can derive a transport equation for the wake velocity deficit, and subsequently the mean wake velocity deficit, by doubly decomposing Eq.~\ref{eqn: LES momentum} as described above. This procedure yields the following transport equation for the mean wake velocity deficit
\begin{align}\label{eqn:wake momentum transport}
    \begin{split}
        \left(\overline{u_j^B} + \overline{\Delta u_j}\right) \pdv{\overline{\Delta u_i}}{x_j} = - \pdv{\overline{\Delta p}}{x_i}  + \frac{\delta_{i3}}{Fr^2} \overline{\Delta \theta}- \frac{2}{Ro} \varepsilon_{ijk} \Omega_j\overline{\Delta u_k} - \pdv{\overline{\Delta \tau_{ij}}}{x_j} - \overline{\Delta u_j} \pdv{\overline{u_i^B}}{x_j}
         + \pdv{}{x_j}\overline{u_i'u_j'}_{\text{wake}},
    \end{split}
\end{align}
where $\overline{u_i'u_j'}_{\text{wake}}=\overline{u_i'u_j'} - \overline{u_i\Bp u_j\Bp}$. As stated in Section~\ref{sec:intro}, we study the momentum forcings that are critical in different ABL stratifications to better understand the physical mechanisms that are relevant to the wake momentum deficit. We focus on the streamwise velocity deficit $\overline{\Delta u}$ to interrogate the assumptions present in models that stem from one-dimensional momentum theory. Focusing on this component, we arrive at a transport equation for the streamwise momentum deficit given by
\begin{eqnarray}\label{eqn:streamwise wake momentum transport}
    \left(\overline{u_j^B} + \overline{\Delta u_j}\right) \pdv{\overline{\Delta u}}{x_j} = - \pdv{\overline{\Delta p}}{x} +\frac{2}{Ro}\Omega_3\overline{\Delta v} - \pdv{\overline{\Delta \tau_{1j}}}{x_j} - \overline{\Delta u_j} \pdv{\overline{u^B}}{x_j}  + \pdv{}{x_j}\overline{ u' u_j'}_{\text{wake}},
\end{eqnarray}
using the traditional approximation for the Coriolis term that only retains the vertical component of earth's rotation~\cite{howland_influence_2020}, which is also enforced in the LES.

Equation~\ref{eqn:streamwise wake momentum transport} can be used to solve for $\overline{\Delta u}$ at any arbitrary $x$ location by rearranging and integrating, such that 
\begin{align}\label{eqn:streamwise wake momentum transport integral}
    \begin{split}
        \overline{\Delta u} = \int_{x_0}^{x} &\frac{1}{\overline{u^B} + \overline{\Delta u}}\left(\underbrace{-\overline{v^B} \pdv{\overline{\Delta u}}{y} -\overline{w^B} \pdv{\overline{\Delta u}}{z}}_{\rm{I}} \underbrace{- \pdv{\overline{\Delta p}}{x} - \pdv{\overline{\Delta \tau_{1j}}}{x_j}  + \frac{2}{Ro}\Omega_3\overline{\Delta v}}_{\rm{II}} \right . 
        \\& \left .\underbrace{- \overline{\Delta v} \pdv{\overline{\Delta u}}{y} - \overline{\Delta w} \pdv{\overline{\Delta u}}{z}- \overline{\Delta v} \pdv{\overline{u^B}}{y} - \overline{\Delta w} \pdv{\overline{u^B}}{z}}_{\rm{III}}  \underbrace{-\pdv{}{x_j}\overline{ u' u_j'}_{\text{wake}}}_{\rm{IV}} \right )dx',
    \end{split}
\end{align}
where the terms have been grouped as follows: I represents advection of the wake velocity deficit by the base flow (termed: $-$Adv $\overline{v^B}, \; \overline{w^B}$), II represents the pressure gradient (Pres), SGS, and Coriolis (Cor) terms, III represents advection by the deficit velocity of both $\overline{\Delta u}$ and $\overline{u^B}$ ($-$Adv $\overline{\Delta v}, \; \overline{\Delta w}$), and IV represents the turbulence divergence (Turb). In Eq.~\ref{eqn:streamwise wake momentum transport integral}, all terms in general depend on $y$ and $z$.

\subsubsection{\label{subsubsec: wake-added tke} Wake-added turbulence kinetic energy}
Turbulence kinetic energy provides information about the energy content of the turbulence in a particular flow and how it is spatially distributed. The TKE is a critical quantity of interest often used to model mean flow in the wake and also affects the loads of downwind waked turbines. However, a challenge in investigating wake-added TKE in the stratified ABL is that the turbulent inflow itself contains TKE that depends on ABL roughness and stratification. Here, we again look at the deficit budget for this quantity to isolate the wake from the incident base flow. Wake-added TKE, denoted by $k_{\text{wake}}$, is defined as 
\begin{eqnarray}
    k_{\text{wake}} = \frac{1}{2}\left (\overline{u_i' u_i'} - \overline{u_i\Bp u_i\Bp} \right ),
\end{eqnarray}
where $\frac{1}{2}\overline{u_i' u_i'}$ is the TKE from the full flow field with the turbine and $\frac{1}{2}\overline{u_i\Bp u_i\Bp}$ is the TKE from the base flow field. A transport equation for $k_{\text{wake}}$ can be derived analogously to that of $\overline{\Delta u}$. Starting from the transport equation for the full flow field TKE $k$ given by
\begin{eqnarray}\label{eqn:tke transport}
    \overline{u_j}\pdv{k}{x_j} = -\overline{u_i' u_j'} \pdv{\overline{u_i}}{x_j} + \frac{1}{ Fr^2} \overline{w' \theta'}- \pdv{}{x_j} \left (\overline{u_j' k} + \overline{u_j' p'} - \overline{\tau_{ij}'u_j'} \right) - \overline{\tau_{ij}' \pdv{u_i'}{x_j}},
\end{eqnarray}
where the terms on the right-hand side are as follows from left to right: shear production (termed: Prod), buoyancy (Buoy), turbulent transport (Turb), pressure transport (Pres), SGS transport (SGS), and dissipation (Diss). Subtracting the transport equation for the base flow TKE $k^B$ from Eq.~\ref{eqn:tke transport} yields the transport equation for $k_{\text{wake}}$. given by
\begin{align}\label{eqn:wake tke transport}
    \begin{split}
    \overline{u_j}\pdv{k_\text{wake}}{x_j}= & -\overline{\Delta u_j}\pdv{k^B}{x_j} -\overline{u_i' u_j'} \pdv{\overline{u_i}}{x_j} +\overline{u_i\Bp u_j\Bp} \pdv{\overline{u_i^B}}{x_j} + \frac{1}{ Fr^2} \left(\overline{w' \theta'} - \overline{w\Bp \theta\Bp} \right ) 
    \\&- \pdv{}{x_j} \left (\overline{u_j' k} - \overline{u_j\Bp k^B} + \overline{u_j' p'} - \overline{u_j\Bp p\Bp} - \overline{\tau_{ij}'u_j'} + \overline{\tau_{ij}\Bp u_j\Bp} \right) 
    - \overline{\tau_{ij}' \pdv{u_i'}{x_j}} +  \overline{\tau_{ij}\Bp \pdv{u_i\Bp}{x_j}},
    \end{split}
\end{align}
where the terms are analogously defined to those in Eq.~\ref{eqn:tke transport} but now govern the transport of wake-added TKE. The first term on the right-hand side is an additional term that represents the advection of the base flow TKE by the wake deficit flow. As in the standard TKE equation, Coriolis effects do not directly appear, but are present indirectly through the modification of the transport and redistribution of momentum and turbulence. Following the analysis in Section~\ref{subsubsec: streamwise momentum}, we can rearrange Eq.~\ref{eqn:wake tke transport} and integrate in the streamwise direction to solve for $k_\mathrm{wake}$:
\begin{align}\label{eqn:wake tke transport integral}
    \begin{split}
     k_\text{wake}= & {\int_{x_0}^x }\frac{1}{\overline{u^B} + \overline{\Delta u}} \left [ 
    \underbrace{-\overline{v^B} \pdv{k_\mathrm{wake}}{y} -\overline{w^B} \pdv{k_\mathrm{wake}}{z}}_{\mathrm{I}_k} \underbrace{-\overline{\Delta v} \pdv{k_\mathrm{wake}}{y} -\overline{\Delta w} \pdv{k_\mathrm{wake}}{z}}_{\mathrm{II}_k} + \underbrace{ \vphantom{-\overline{v^B} \pdv{k}{y} -\overline{w^B} \pdv{k}{z}}\frac{1}{ Fr^2} \left(\overline{w' \theta'} - \overline{w\Bp \theta\Bp} \right )}_{\mathrm{III}_k}
    \right .
    \\&\left .\underbrace{-\overline{\Delta u_j}\pdv{k^B}{x_j} - \pdv{}{x_j} \left (\overline{u_j' p'} - \overline{u_j\Bp p\Bp} - \overline{\tau_{ij}'u_j'} + \overline{\tau_{ij}\Bp u_j\Bp} \right)}_{\mathrm{IV}_k} \right .
    \\ & \left .
    \underbrace{- \overline{\tau_{ij}' \pdv{u_i'}{x_j}} +  \overline{\tau_{ij}\Bp \pdv{u_i\Bp}{x_j}}}_{\mathrm{V}_k} \underbrace{- \pdv{}{x_j} \left (\overline{u_j' k} - \overline{u_j\Bp k^B}\right )}_{\mathrm{VI}_k} \underbrace{-\overline{u_i' u_j'} \pdv{\overline{u_i}}{x_j} +\overline{u_i\Bp u_j\Bp} \pdv{\overline{u_i^B}}{x_j}}_{\mathrm{VII}_k}\right ] dx'
    \end{split}
\end{align}
The terms in Eq.~\ref{eqn:wake tke transport integral} above are as follows: I$_k$ is advection of $k_\mathrm{wake}$ by the base flow; II$_k$ is advection of $k_\mathrm{wake}$ by the deficit flow; III$_k$ is buoyant production/destruction; IV$_k$ is a combination of advection of $k^B$ by the deficit flow, transport of $k_\mathrm{wake}$ by pressure fluctuations, and transport of $k_\mathrm{wake}$ by SGS fluctuations; V$_k$ is dissipation; VI$_k$ is turbulent transport of $k_\mathrm{wake}$; and VII$_k$ is shear production. These groupings will be utilized in Section~\ref{subsec:results tke}.

\subsection{\textit{A priori} analysis}\label{subsec:a priori}

As stated in the preceding sections (Section~\ref{subsubsec: streamwise momentum}-\ref{subsubsec: wake-added tke}), the steady state transport equations for $\overline{\Delta u}$ (Eq.~\ref{eqn:streamwise wake momentum transport integral}) and $k_\mathrm{wake}$ (Eq.~\ref{eqn:wake tke transport integral}) can be used to solve for these variables by integrating in the streamwise direction. We exploit this fact to analyze the importance of each physical mechanism in both the wake velocity deficit and wake-added TKE budgets, through a forward marching \textit{a priori} model analysis using LES data, in which terms (or groups of terms) are removed one at a time. By removing a given physical mechanism and then marching the solution forward in space, we can ascertain the impact of said mechanism, which informs both our understanding of the differences between differently stratified flows and the modeling for such flows.

\subsection{Control volume analysis}\label{subsec:control volume}

To analyze the integral budgets for $\overline{\Delta u}$, we employ two different control volume analyses. In the first, the control volume is denoted by the streamtube enclosing the wake. The streamtube control volume analysis is motivated by the long history of one-dimensional momentum theory~\cite{sorensen_general_2016}, which informs present analytical wake modeling tools. By analyzing changes within a streamtube, we study what is important for wake recovery. 

We also use a box control volume as is done in other works~\cite{cortina_distribution_2016,cortina_wind_2017,cortina_mean_2020}. By using a box that encompasses the streamtube volume and also part of the surrounding flow, we can study effects that are not captured by the streamtube analysis. Specifically, it is possible to learn about the forcings that alter the shape of the wake, which  is particularly important in flows with high degrees of speed and direction shear. Together these two control volume analyses aid in our understanding of what is important within the wake as previously studied in momentum theory (streamtube) and which mechanisms acting outside the wake streamtube affect the wake deficit (box). 

\section{\label{sec:results} Results}
For the analysis, we focus on the streamwise momentum deficit and wake-added TKE budgets. The streamwise momentum deficit budget offers insight into the physical mechanisms that are most important in the wake and--as a result--what is important in modeling $\overline{\Delta u}$. As such, it is instructive to see how this quantity is affected as stratification changes and how this can inform modeling efforts. We also analyze the wake-added TKE budget to study the character of the turbulence in these two stratified ABL flows. Wake-added TKE or wake-added turbulence intensity is also often modeled in wake models. Here, we look at the budgets of these two quantities directly to isolate their dynamics from the base flow dynamics. Together, these two physical quantities and the analysis of the associated budgets provide insight into the physical mechanisms relevant in the wake as stratification changes for both the wake deficit and the turbulence. 

\subsection{\label{subsec:methods abl simulations} Atmospheric boundary layer comparison}
\begin{figure}
    \centering
    \subfloat[Wind speed\label{fig:ws comp}]{
        \includegraphics[width=0.45\textwidth]{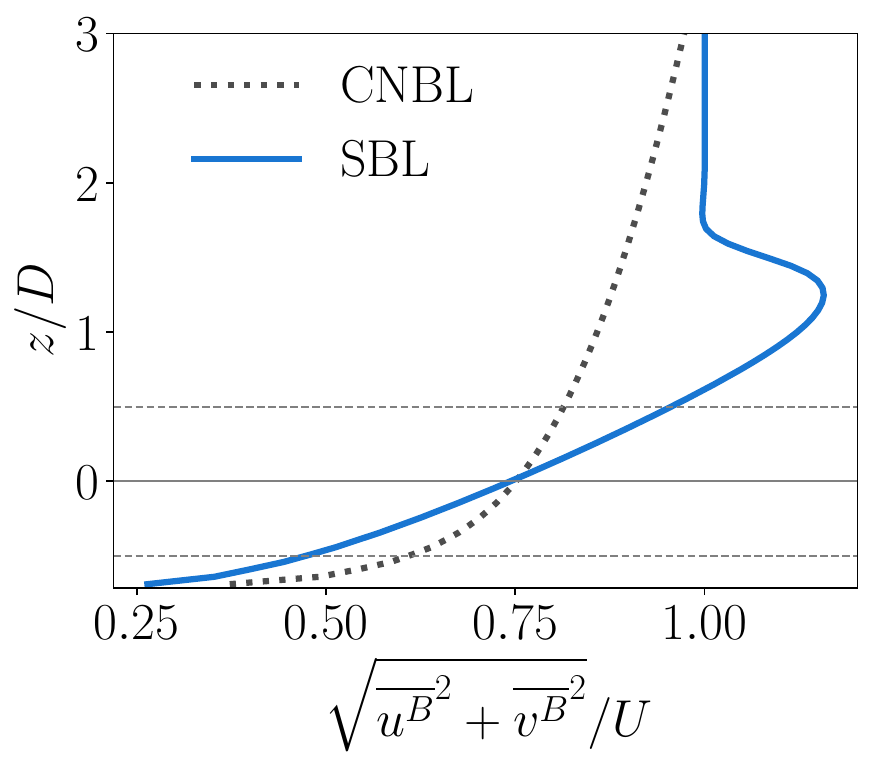}}
    \subfloat[Potential temperature\label{fig:pot T comp}]{\includegraphics[width=0.45\textwidth]{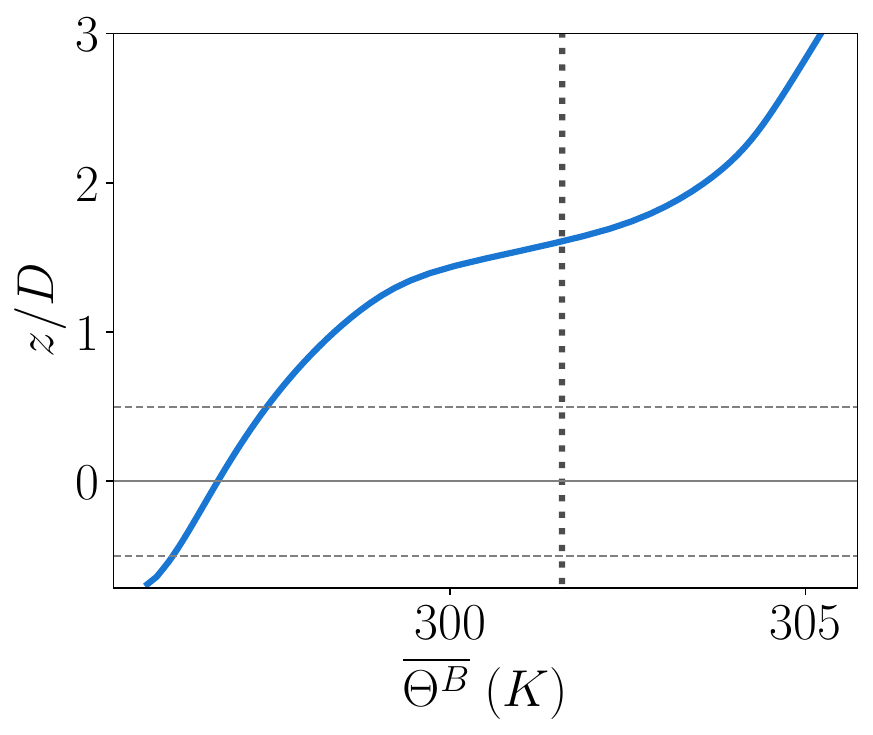}}\\
    \subfloat[Wind direction\label{fig:wd comp}]{
        \includegraphics[width=0.45\textwidth]{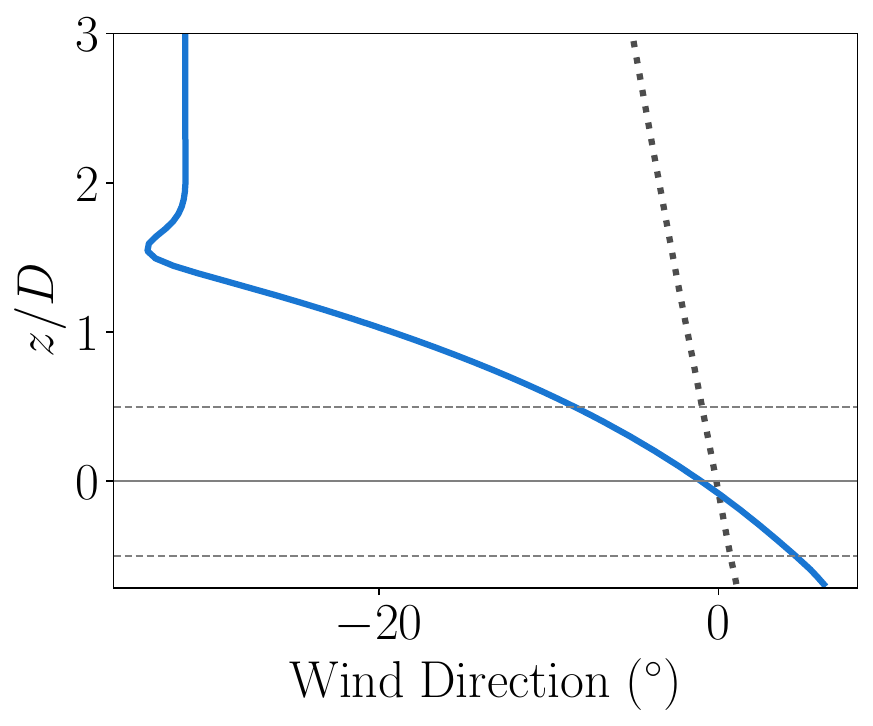}}
    \subfloat[TKE\label{fig:tke comp}]{
            \includegraphics[width=0.45\textwidth]{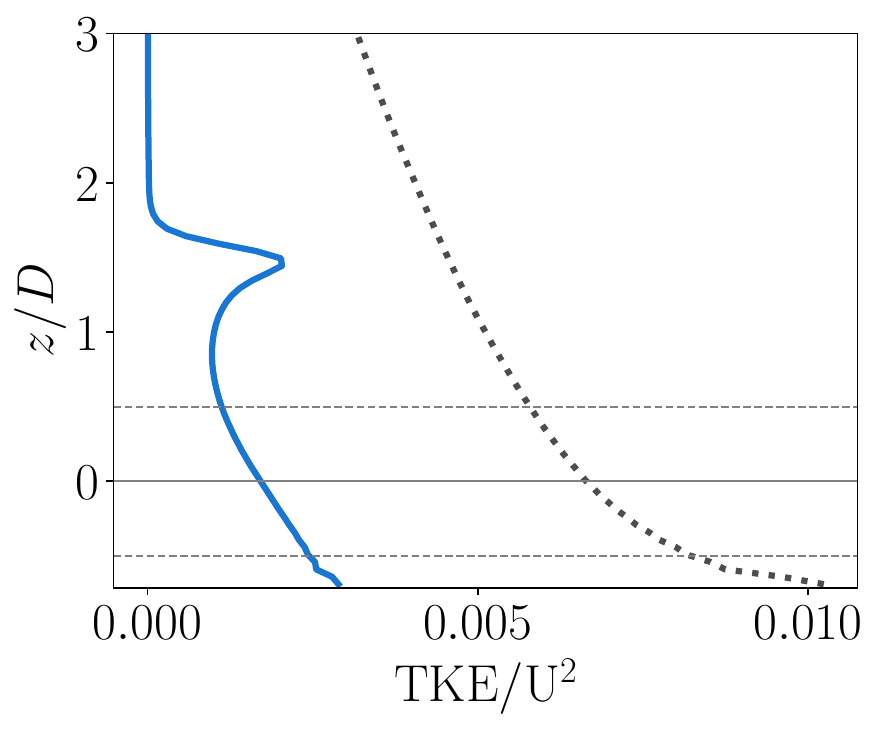}}
    \caption{Comparison of base flow quantities in the CNBL and SBL. All quantities are both time-averaged and spatially averaged in the streamwise and lateral directions. Wind speed and TKE are normalized via the geostrophic wind speed $U$. Note that this is a subset of the vertical domain, zoomed in near the turbine.}\label{fig:nbl sbl comp}
\end{figure}

In comparing the base CNBL and SBL inflow conditions, there are a number of important differences. Figure~\ref{fig:nbl sbl comp} shows a comparison of the wind speed, wind direction, potential temperature, and TKE profiles for the CNBL and SBL. The SBL exhibits much higher velocity shear and veer than the CNBL, while the CNBL has higher turbulence content as indicated by the TKE profiles. In all figures with mean quantities in this work, the quantities are time-averaged. Figures are also spatially averaged where indicated.

\begin{figure}
    \centering
    \subfloat[Wake velocity deficit\label{fig:delta u comp xy}]{
        \includegraphics[width=0.45\textwidth]{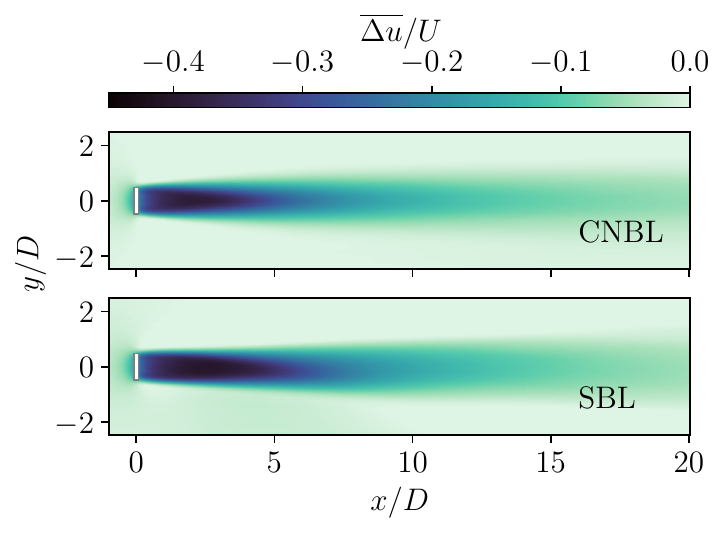}}
    \subfloat[Wake-added TKE\label{fig:wake tke comp xy}]{
        \includegraphics[width=0.45\textwidth]{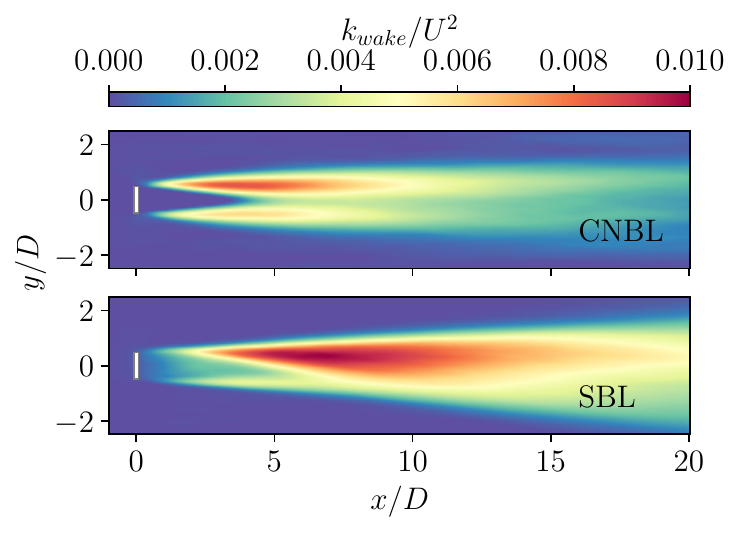}}
    \caption{Comparison of $\overline{\Delta u}$ and $k_\mathrm{wake}$ in the $x-y$ plane at hub height.}\label{fig:wake tke delta u comp xy}
\end{figure}

Figure~\ref{fig:wake tke delta u comp xy} shows the two quantities of interest in this work in the $x-y$ plane: $\overline{\Delta u}$ (Fig.~\ref{fig:delta u comp xy}) and $k_\mathrm{wake}$ (Fig.~\ref{fig:wake tke delta u comp xy}). The core of the SBL wake velocity deficit is longer than that of the CNBL (as is expected~\cite{abkar_influence_2015}). We note that in both the CNBL and SBL, $k_\mathrm{wake}$ exhibits asymmetric behavior. This behavior is largely due to the wind direction changes with height that are induced by Coriolis forcing. These wind direction changes are much larger in the SBL (see Fig.~\ref{fig:wd comp}), which leads to more pronounced asymmetry. This asymmetry in $k_\mathrm{wake}$ is observed in Wu et al.~\cite{wu_new_2023}. Interestingly, $k_\mathrm{wake}$ also appears to be higher in the SBL than in the CNBL, which will be discussed further in Section~\ref{subsec:discussion}.

\subsection{\label{subsec:results momentum} Streamwise wake momentum budget}
As stated above, the streamwise wake momentum budget is studied here to assess the importance of the physical mechanisms in the wake for flows in different stability regimes (conventionally neutral and stable). This is accomplished in two ways. First, in Section~\ref{subsubsec:results momentum a priori} we use Eq.~\ref{eqn:streamwise wake momentum transport integral} to solve for $\overline{\Delta u}$ using LES data in an \textit{a priori} manner as outlined in Section~\ref{subsec:a priori}, systematically removing terms. Then in Section~\ref{subsubsec:results momentum budget}, we use streamtube and control volume budget analysis to complement the \textit{a priori} analysis and illustrate how streamtube and control volume integration in the wake provide two viewpoints for assessing the processes in the wake streamtube that govern wake recovery and the physics that affect the overall wake shape.

\subsubsection{\label{subsubsec:results momentum a priori}\textit{A priori} wake analysis}
\begin{figure*}
    \centering
    \includegraphics[width=0.95\textwidth]{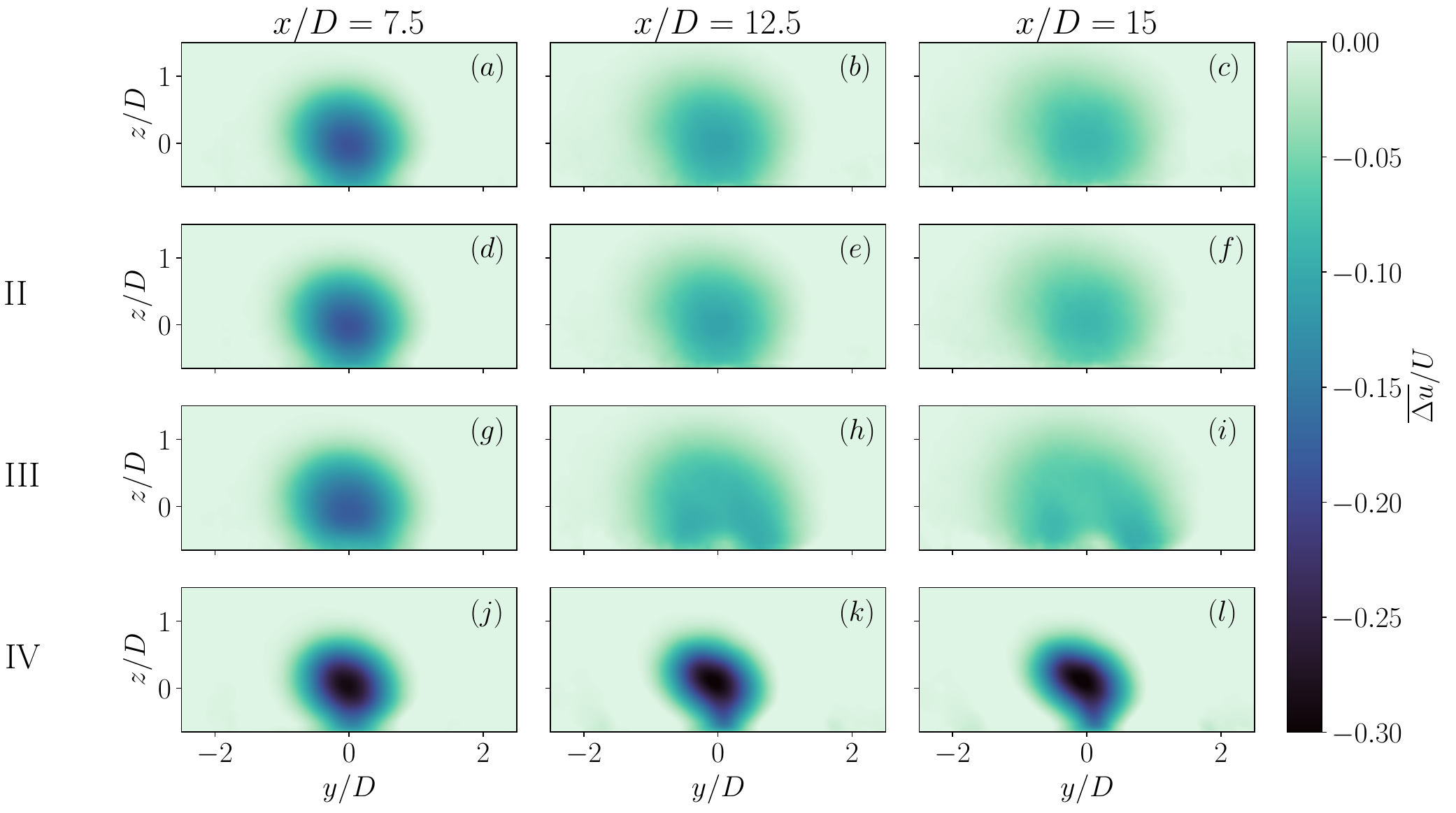}
    \caption{CNBL velocity deficit for \textit{a priori} evaluation of the momentum budget. The leftmost column figures ($a$, $d$, $g$, $j$) are $y-z$ slices of the wake taken at $x/D=7.5$, the middle column figures ($b$, $e$, $h$, $k$) are taken at $x/D=12.5$, and the rightmost column figures ($c$, $f$, $i$, $l$) are taken at $x/D=15$. From top to bottom: first row figures ($a$-$c$) correspond to $\overline{\Delta u}/U$ computed with no terms removed, second row figures ($d$-$f$) (II) correspond to $\overline{\Delta u}/U$ computed with pressure gradient, SGS, and Coriolis terms removed, third row figures ($g$-$i$) (III) correspond to $\overline{\Delta u}/U$ computed with the advection terms that depend on $\overline{\Delta v}$ and $\overline{\Delta w}$ removed, and fourth row figures ($j$-$l$) (IV) correspond to $\overline{\Delta u}/U$ computed with turbulence terms removed.}\label{fig:nbl delta u grid physics}
\end{figure*}

\begin{figure*}
    \centering
    \includegraphics[width=0.95\textwidth]{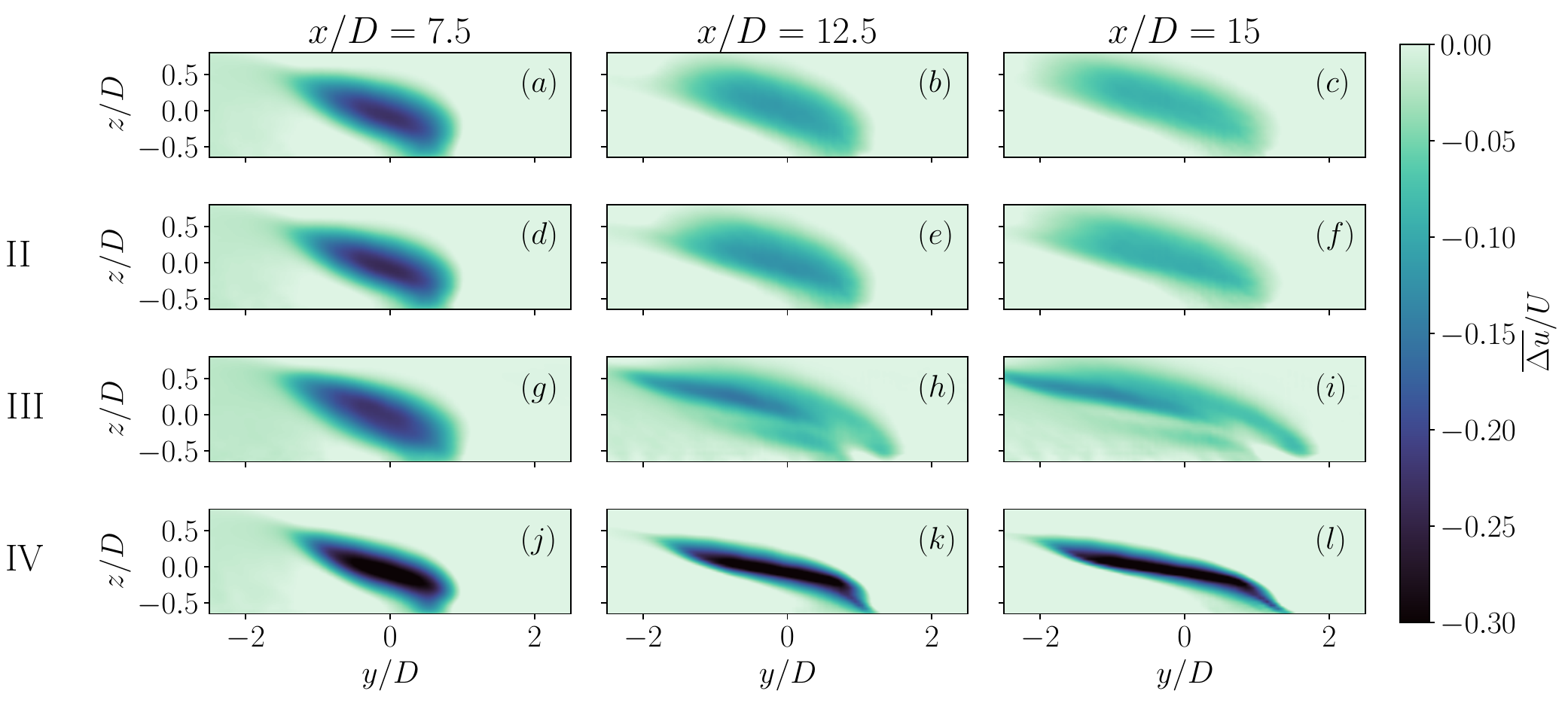}
    \caption{SBL velocity deficit for \textit{a priori} evaluation of the momentum budget. The leftmost column figures ($a$, $d$, $g$, $j$) are $y-z$ slices of the wake taken at $x/D=7.5$, the middle column figures ($b$, $e$, $h$, $k$) are taken at $x/D=12.5$, and the rightmost column figures ($c$, $f$, $i$, $l$) are taken at $x/D=15$. From top to bottom: first row figures ($a$-$c$) correspond to $\overline{\Delta u}/U$ computed with no terms removed, second row figures ($d$-$f$) (II) correspond to $\overline{\Delta u}/U$ computed with pressure gradient, SGS, and Coriolis terms removed, third row figures ($g$-$i$) (III) correspond to $\overline{\Delta u}/U$ computed with the advection terms that depend on $\overline{\Delta v}$ and $\overline{\Delta w}$ removed, and fourth row figures ($j$-$l$) (IV) correspond to $\overline{\Delta u}/U$ computed with turbulence terms removed.}\label{fig:sbl delta u grid physics}
\end{figure*}

For the \textit{a priori} analysis, we solve for $\overline{\Delta u}$ via Eq.~\ref{eqn:streamwise wake momentum transport integral} using forcing terms calculated using LES data. In order to isolate the importance of different physical mechanisms, terms (or groups of terms) are removed one at time. This analysis is shown in Figs.~\ref{fig:nbl delta u grid physics} (CNBL) and~\ref{fig:sbl delta u grid physics} (SBL). Focusing on $\overline{\Delta u}/U$ in the CNBL in Fig.~\ref{fig:nbl delta u grid physics}, from top to bottom the first row has no terms removed, the second row has the pressure gradient, SGS, and Coriolis terms removed, the third row has $\overline{\Delta v}$ and $\overline{\Delta w}$ removed, and the bottom row has the turbulence term removed. The three different columns are different streamwise locations, where the leftmost column is at $x/D=7.5$, the middle column is at $x/D=12.5$, and the rightmost column is at $x/D=15$. Note that term I in Eq.~\ref{eqn:streamwise wake momentum transport integral} (advection of $\overline{\Delta u}$ by the base flow lateral and vertical velocities) is not removed in this analysis. This choice is made in part based on the form of the curled wake model~\cite{martinez-tossas_curled_2021}, in which the base flow is given as an input to the model. In this \textit{a priori} analysis, we consider terms that are neglected or fully modeled in the curled wake model. In a general setting, we are interested in the modeling context where information about the base ABL flow (inflow) is known, and we seek to predict the wake flow that depends on the inflow condition.

In analyzing the wake velocity deficits in Figs.~\ref{fig:nbl delta u grid physics} and~\ref{fig:sbl delta u grid physics}, it is instructive to compare each row where terms have been removed to the top row of figures in which all the terms from the LES have been used to compute $\overline{\Delta u}/U$. In doing so, it is clear that the largest differences arise when the turbulence is removed from the computation (Figs.~\ref{fig:nbl delta u grid physics}($j$)-($l$) and~\ref{fig:sbl delta u grid physics}($j$)-($l$)). It should also be noted that no turbulence model has been utilized in place of the turbulence divergence term calculated from the LES. Of particular note are the magnitude and shape of the wake deficit when the turbulence is absent. The magnitude of the deficit is larger and the wake is much more confined in the absence of turbulent diffusion, indicating the importance of turbulent mixing in the wake. 

Of next highest importance are the terms advected by the lateral and vertical wake deficit velocities, particularly for the SBL. Unlike in the case where the turbulence has been removed, when the $\overline{\Delta v}$ and $\overline{\Delta w}$ advection terms are removed it is primarily the shape that is affected and not the magnitude of the velocity deficit. This is because these terms are related to the veering or the skewing of the wake deficit, so their omission leads to an incorrect wake shape. Again, this is seen most strongly in the SBL case, where veering is more prominent (see Fig.~\ref{fig:wd comp}). 

Finally, we find that the effect of removing the pressure gradient, SGS, and Coriolis terms (Figs.~\ref{fig:nbl delta u grid physics}($d$)-($f$) and~\ref{fig:sbl delta u grid physics}($d$)-($f$)) is relatively negligible for these ABL conditions. Between the two ABL conditions, there appears to be a slightly larger effect on the shape of the SBL wake deficit.

\begin{figure}
    \centering
    \includegraphics[width=0.95\textwidth]{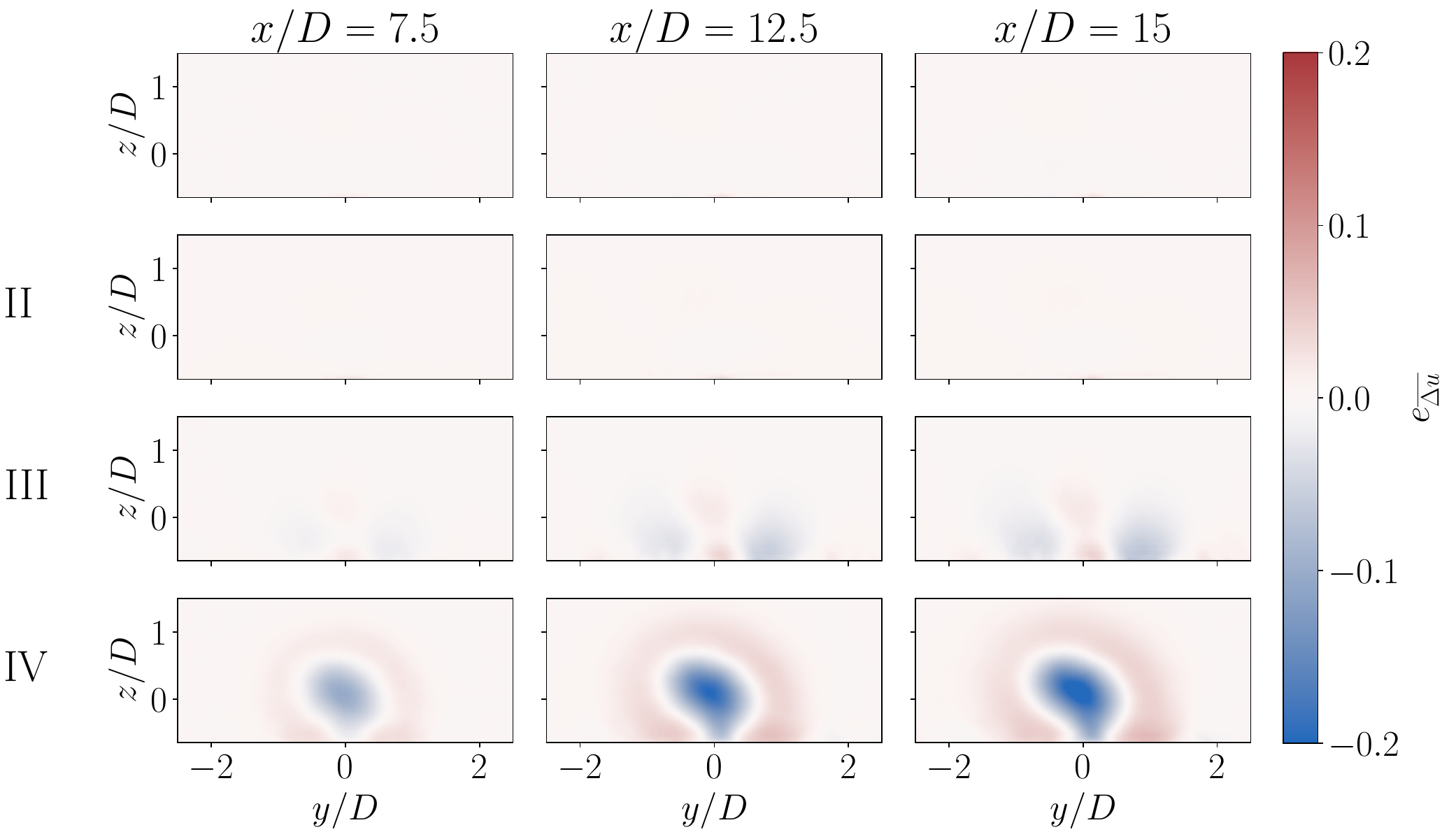}
        \caption{CNBL velocity deficit error as defined by Eq.~\ref{eq:delta u error}. The leftmost column figures ($a$, $d$, $g$, $j$) are $y-z$ slices of the wake taken at $x/D=7.5$, the middle column figures ($b$, $e$, $h$, $k$) are taken at $x/D=12.5$, and the rightmost column figures ($c$, $f$, $i$, $l$) are taken at $x/D=15$. From top to bottom: first row figures ($a$-$c$) correspond to $\overline{\Delta u}/U$ computed with no terms removed, second row figures ($d$-$f$) (II) correspond to $\overline{\Delta u}/U$ computed with pressure gradient, SGS, and Coriolis terms removed, third row figures ($g$-$i$) (III) correspond to $\overline{\Delta u}/U$ computed with the advection terms that depend on $\overline{\Delta v}$ and $\overline{\Delta w}$ removed, and fourth row figures ($j$-$l$) (IV) correspond to $\overline{\Delta u}/U$ computed with turbulence terms removed.}\label{fig:nbl delta u error grid physics}
\end{figure}

\begin{figure}
    \centering
    \includegraphics[width=0.95\textwidth]{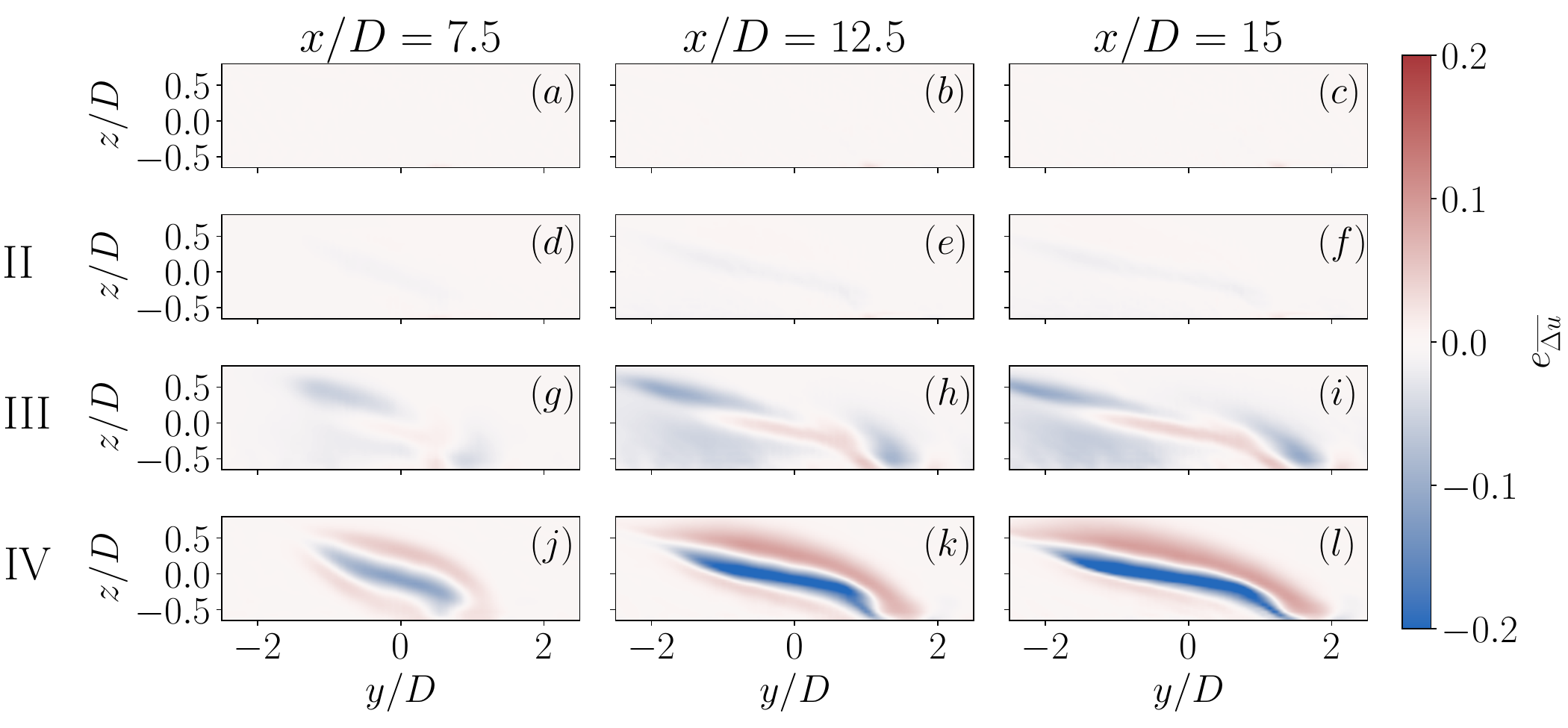}
    \caption{SBL velocity deficit error as defined by Eq.~\ref{eq:delta u error}. The leftmost column figures ($a$, $d$, $g$, $j$) are $y-z$ slices of the wake taken at $x/D=7.5$, the middle column figures ($b$, $e$, $h$, $k$) are taken at $x/D=12.5$, and the rightmost column figures ($c$, $f$, $i$, $l$) are taken at $x/D=15$. From top to bottom: first row figures ($a$-$c$) correspond to $\overline{\Delta u}/U$ computed with no terms removed, second row figures ($d$-$f$) (II) correspond to $\overline{\Delta u}/U$ computed with pressure gradient, SGS, and Coriolis terms removed, third row figures ($g$-$i$) (III) correspond to $\overline{\Delta u}/U$ computed with the advection terms that depend on $\overline{\Delta v}$ and $\overline{\Delta w}$ removed, and fourth row figures ($j$-$l$) (IV) correspond to $\overline{\Delta u}/U$ computed with turbulence terms removed.}\label{fig:sbl delta u error grid physics}
\end{figure}

A quantitative comparison of the importance of the various physical mechanisms comes from taking the difference between the velocity deficit from the \textit{a priori} analysis $\overline{\Delta u}_{\textit{a priori}}$ and the velocity deficit from LES $\overline{\Delta u}_{\text{LES}}$:
\begin{eqnarray}\label{eq:delta u error}
    e_{\overline{\Delta u}} = \frac{\overline{\Delta u}_{\textit{a priori}} - \overline{\Delta u}_{\text{LES}}}{U}.
\end{eqnarray}
Figures~\ref{fig:nbl delta u error grid physics} and~\ref{fig:sbl delta u error grid physics} show the error defined in Eq.~\ref{eq:delta u error}, where the columns correspond to the same $x/D$ locations and the rows correspond to the same flow physics as in Figs.~\ref{fig:nbl delta u grid physics} and~\ref{fig:sbl delta u grid physics}. The error reveals what was found in the qualitative analysis of Figs.~\ref{fig:nbl delta u grid physics} and~\ref{fig:sbl delta u grid physics}: removing the turbulent diffusion results in the largest errors, removing $\overline{\Delta v}$ and $\overline{\Delta w}$ advection results in secondary errors, and the error from removing the pressure gradient, SGS, and Coriolis terms is relatively negligible. In both the CNBL and SBL, the error grows for all cases moving downstream. Additionally, the SBL has larger errors for both the removal of $\overline{\Delta v}$ and $\overline{\Delta w}$ advection and the removal of the pressure gradient, SGS, and Coriolis terms. 

\begin{figure}
    \centering
    \begin{subfigure}{0.4\textwidth}
        \includegraphics[width=\textwidth]{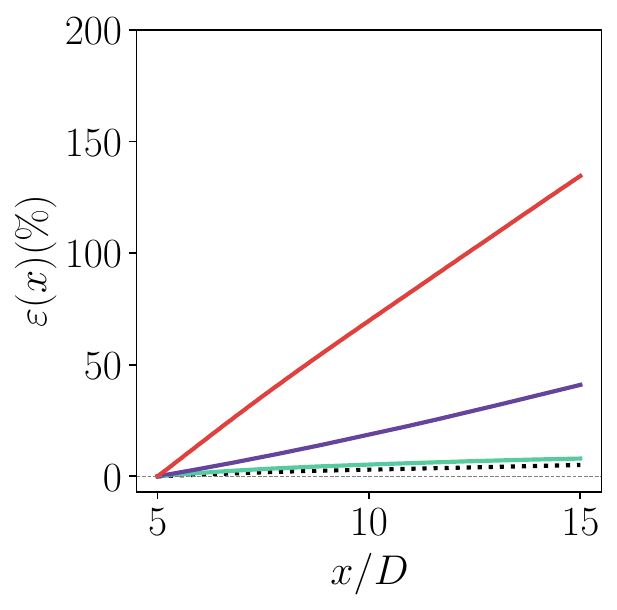}
        \caption{CNBL}
    \end{subfigure}
    \begin{subfigure}{0.4\textwidth}
        \includegraphics[width=\textwidth]{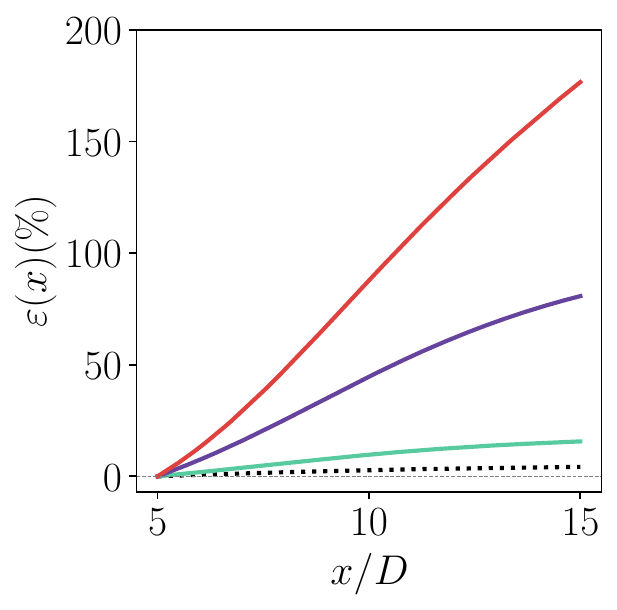}
        \caption{SBL}
    \end{subfigure}
    \caption{Velocity deficit error in $x$ as defined in Eq.~\ref{eq:delta u error in x} for SGS, Coriolis, and pressure gradient removed (II) (\errnodpdxsgscor[scale=0.5]), $\overline{\Delta v}$ and $\overline{\Delta w}$ advection terms removed (III) (\errnodeltavw[scale=0.5]), turbulence divergence removed (IV) (\errnoRSdiv[scale=0.5]), and no terms removed (\errfullLES[scale=0.5]). The dashed gray line shows the location of zero error.}\label{fig:nbl sbl error in x}
\end{figure}

We can additionally analyze the error as it accumulates moving downstream by taking the $\ell^2$ norm of the error in Eq.~\ref{eq:delta u error} and normalizing by the $\ell^2$ norm of $\overline{\Delta u}_{\text{LES}}$, which results in
\begin{eqnarray}\label{eq:delta u error in x}
    \varepsilon(x) = 100\% \times \frac{\lVert\overline{\Delta u}_{a\; priori}(x) - \overline{\Delta u}_\mathrm{LES}(x)\rVert_2}{\lVert \overline{\Delta u}_\mathrm{LES}(x)\rVert_2}.
\end{eqnarray}
This error metric is shown in Fig.~\ref{fig:nbl sbl error in x}. For both the CNBL and SBL, the error grows with downstream distance and the relative importance of each term is the same as is found in Figs~\ref{fig:nbl delta u grid physics},~\ref{fig:sbl delta u grid physics},~\ref{fig:nbl delta u error grid physics}, and~\ref{fig:sbl delta u error grid physics}, where removal of turbulence introduces the most error, removal of advection by $\overline{\Delta v}$ and $\overline{\Delta w}$ introduces additional error, and removal of pressure gradient, Coriolis, and SGS introduces fairly negligible error in the CNBL and slightly larger error in the SBL.
\subsubsection{\label{subsubsec:results momentum budget} Streamtube and Control Volume Budgets}

The \textit{a priori} analysis in Section~\ref{subsubsec:results momentum a priori} illustrates which terms are important for the calculation of the streamwise velocity deficit. In this section, we analyze the streamwise momentum deficit budget to complement the \textit{a priori} analysis. We use the terms according to Eq.~\ref{eqn:streamwise wake momentum transport integral}, treating the pressure gradient, SGS, and Coriolis terms separately. We average or integrate within both the wake streamtube and a control volume box enclosing the wake as outlined in Section~\ref{subsec:control volume}. The purpose of employing both these techniques is to study what is important in the wake enclosed by the streamtube---as is done in momentum theory~\cite{rankine_mechanical_1865,froude_elementary_1878,froude_part_1889}---and what affects the wake shape through the box. Multiple studies have used a box as the control volume for studying integrated quantities, such as recovery or mean kinetic energy and harvested power~\cite{cortina_distribution_2016,cortina_wind_2017,cortina_mean_2020}. The control volume box analysis can also be used within a wind farm to locally characterize integrated quantities in the vicinity of individual turbines~\cite{cortina_distribution_2016,cortina_wind_2017,cortina_mean_2020}. Here we use this method around the far wake region of a single turbine.

\begin{figure}
    \centering
    \includegraphics[width=0.95\textwidth]{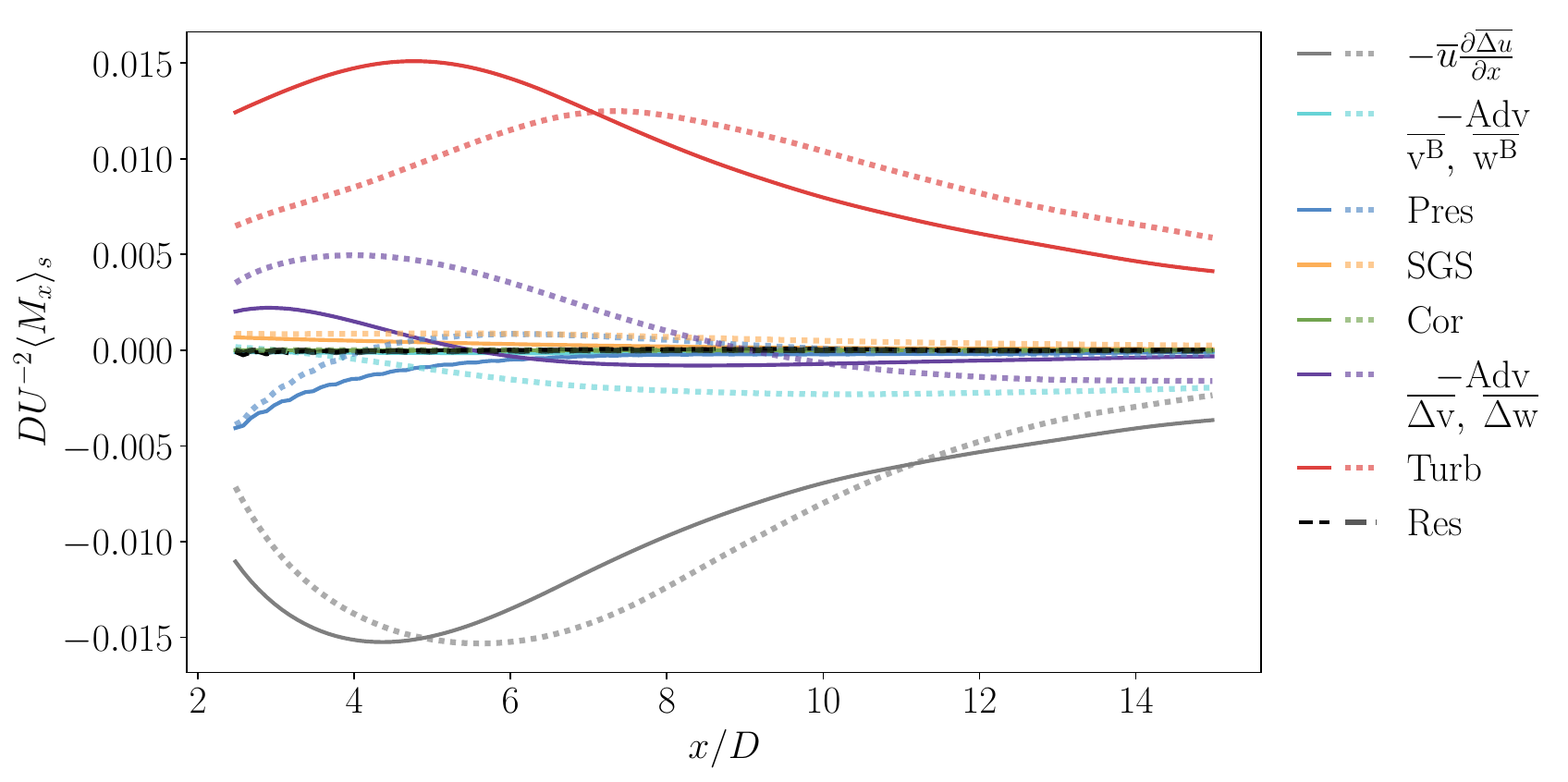}
    \caption{Streamtube-averaged streamwise momentum deficit budget in $x$ for the CNBL (solid lines) and the SBL (dotted lines). Res denotes the budget residual.}\label{fig:nbl sbl xmom streamtube in x}
\end{figure}

Beginning with the streamtube analysis, Figure~\ref{fig:nbl sbl xmom streamtube in x} shows the streamwise momentum deficit budget for both the CNBL and SBL averaged in the streamtube (indicated by $\langle \cdot \rangle_{s}$) at each $x$ location in the wake. For both flows, there is a balance between the mean streamwise advection of the velocity deficit and the turbulence. In the near wake region between $x/D=2.5$ and $x/D=5$, the streamwise pressure gradient and advection by $\overline{\Delta v}$ and $\overline{\Delta w}$ have secondary significance, with this latter term the most significant for the SBL. In the far wake region after $x/D=5$, the balance for the CNBL is again between the mean streamwise advection and the turbulence, with only very minor contributions from advection by $\overline{\Delta v}$ and $\overline{\Delta w}$. For the SBL, there is a more complicated balance; between 5 and 15 diameters downstream, the mean streamwise advection is dominant and balanced by the turbulence with minor effects from advection by both base and deficit lateral and vertical velocities. Moving downstream between roughly 10 and 15 diameters, the lateral and vertical advective terms (both deficit and base) remain relatively constant. Within this region, the mean streamwise advection decreases in magnitude until around $x/D=15$, where all three advection terms are of roughly the same magnitude and together balance the turbulence. Overall, this shows that due to the more dramatic changes in the wind speed and  direction as a function of height $z$ (see Figs.~\ref{fig:ws comp} and~\ref{fig:wd comp}), the lateral and vertical velocities in the wake are more pronounced in the SBL as compared with the CNBL, which reflects the results in the \textit{a priori} analysis in Section~\ref{subsubsec:results momentum a priori}.

\begin{figure}
    \centering
    \includegraphics[width=0.95\textwidth]{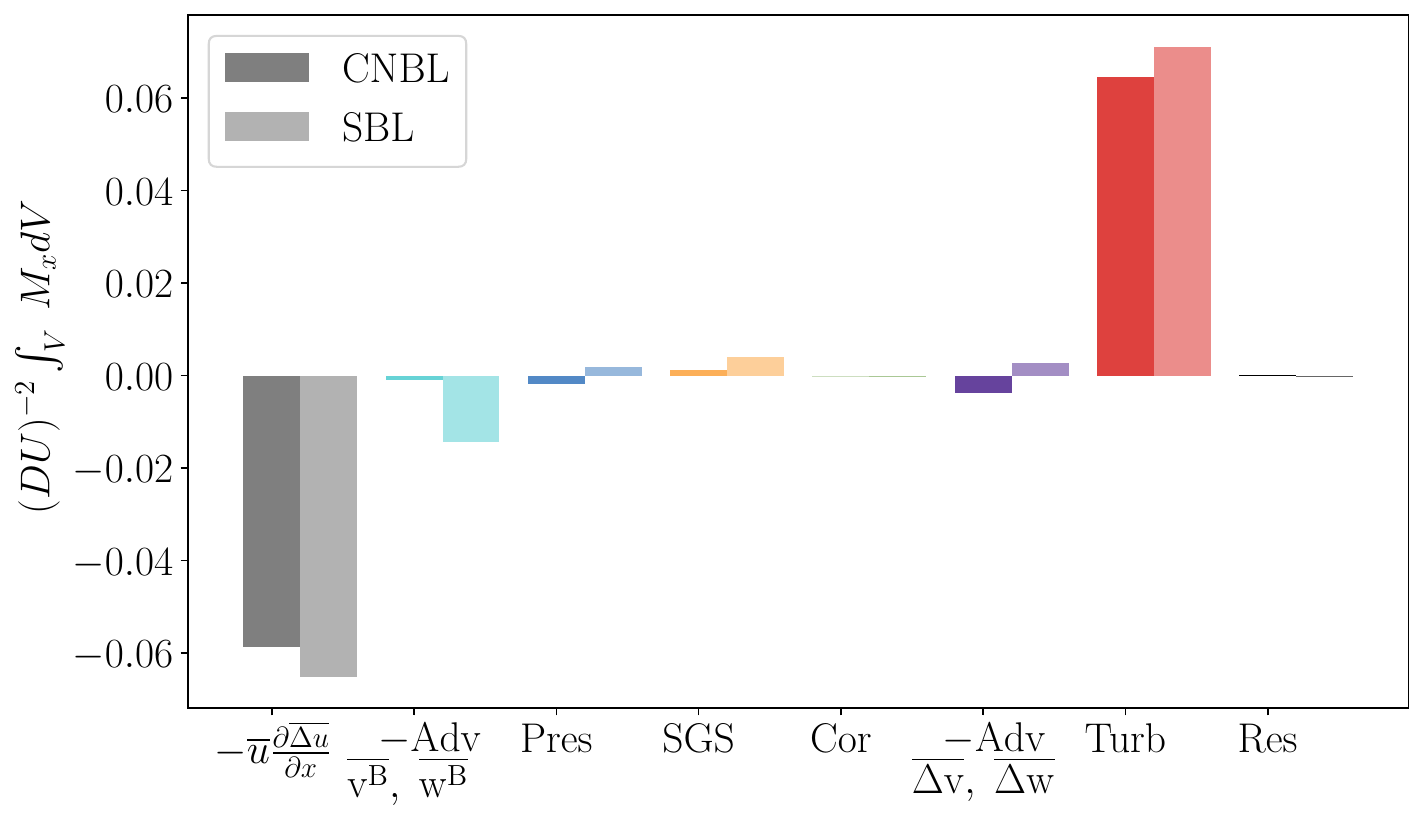}
    \caption{Streamwise momentum deficit budget integrated in the streamtube volume in the far wake ($x/D=5$ to $x/D=15$). Res denotes the budget residual.}\label{fig:nbl sbl xmom streamtube}
\end{figure}

To analyze the overall impact of each term in the streamtube, we further present the streamtube-integrated budget for the streamwise deficit momentum in Figure~\ref{fig:nbl sbl xmom streamtube}. The terms shown are integrated in the streamtube from $x/D=5$ to $x/D=15$ and they show that as with Figure~\ref{fig:nbl sbl xmom streamtube in x} the main balance is between the mean streamwise advection (gray) and the turbulence (red). This view shows how minor the other terms are, particularly for the CNBL. Curiously for the SBL, while the $\overline{v^B}$ and $\overline{w^B}$ advection terms have secondary importance, the $\overline{\Delta v}$ and $\overline{\Delta w}$ advection---which are shown to be significant in Section~\ref{subsubsec:results momentum a priori}---are found to be relatively negligible. Focusing on the streamtube allows a direct evaluation of the wake recovery. From this analysis, we see that the wake recovery is largely linked to the divergence of the Reynolds stresses, which is well-known in and consistent with the literature~\cite{stevens_flow_2017,vanderlaan_brief_2023}. For this reason, modeling approaches seeking to predict the time and streamtube averaged velocity deficit generally neglect other terms, and focus on the parameterization of the Reynolds stress divergence~\cite{shapiro_modelling_2018}.

\begin{figure}
    \centering
    \subfloat[CNBL]{
    \includegraphics[height=0.35\textwidth]{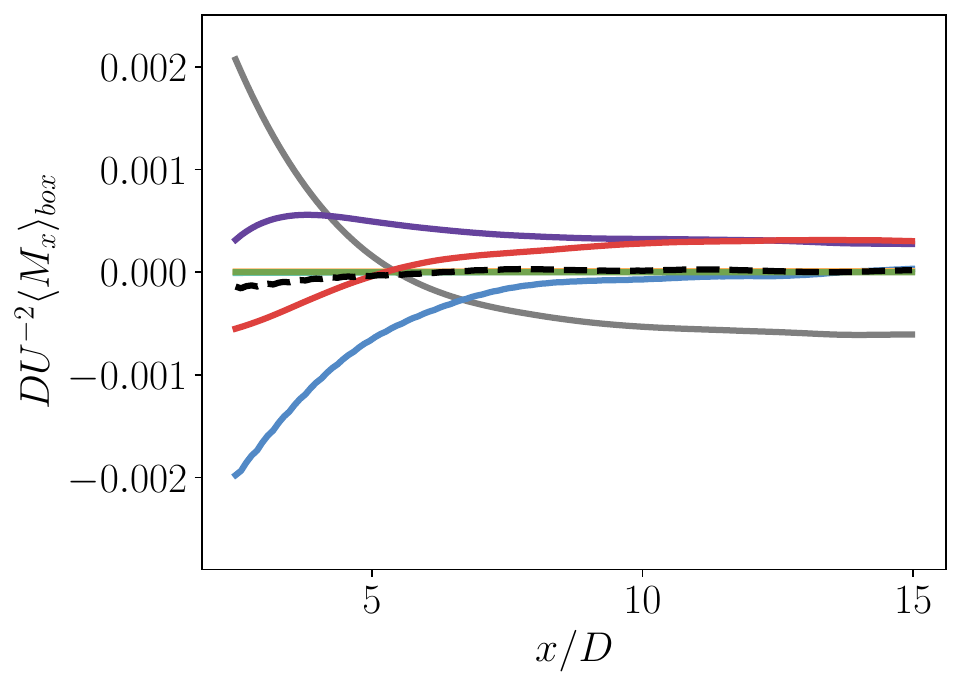}}
    \subfloat[SBL]{
    \includegraphics[height=0.35\textwidth]{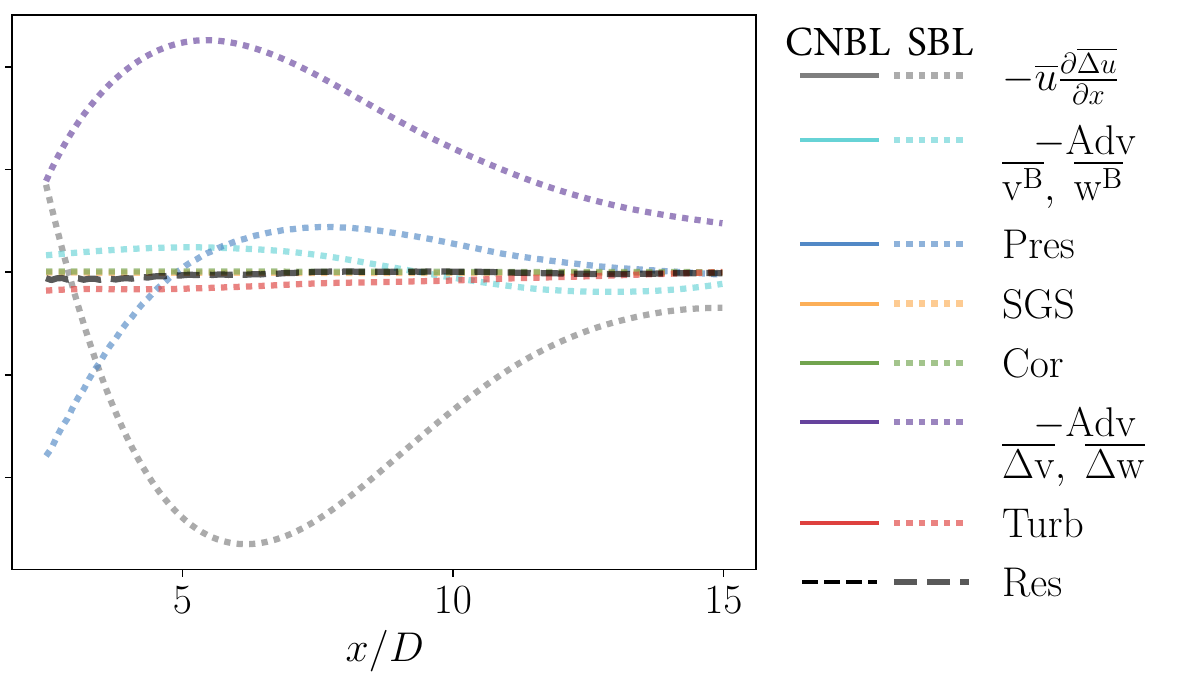}}
    \caption{Box control volume-averaged streamwise momentum deficit budget in $x$ for the CNBL and the SBL. Res denotes the budget residual.}\label{fig:nbl sbl xmom cv in x}
\end{figure}

To further investigate why advection by the lateral and vertical deficit velocities is negligible in the streamtube analysis, we turn to an alternative perspective, now performing the averaging and integration in a box of size $(L_x^{box}, L_y^{box}, L_z^{box}) = (10D, 5D, 1.71D)$ centered on the far wake, such that the dimensions go from $5D$ to $15D$ in the streamwise direction, $-2.5D$ to $2.5D$ in the lateral direction, and $-z_h$ to $z_h + D$ in the vertical direction. In contrast to the streamtube averaging, the box control volume average also quantifies the processes responsible for modifying the wake shape. Figure~\ref{fig:nbl sbl xmom cv in x} shows the box control volume averaged $\langle \cdot \rangle_{box}$ budgets in $x$. For clarity, the CNBL and SBL budgets have been separated into two separate figures. For the CNBL, the balance in the far wake region is primarily between the streamwise advection and the combination of turbulence and $\overline{\Delta v}$ and $\overline{\Delta w}$ advection. This is similar to the streamtube analysis but with increased importance of the deficit vertical and lateral advection terms. For the SBL, we observed something quite different from the streamtube analysis. In both the near and far wake regions of the SBL, the balance is largely between the streamwise advection and $\overline{\Delta v}$ and $\overline{\Delta w}$ advection. There is a large negative peak in the streamwise advection around $x/D=6$, which is not observed in the CNBL. Additionally, the peak in the $\overline{\Delta v}$ and $\overline{\Delta w}$ advection terms in the SBL is about 5 times greater than that of the CNBL. For both the CNBL and SBL there are also secondary contributions from the streamwise pressure gradient, particularly in the near wake region. While they have similar magnitudes, the CNBL streamwise pressure gradient is negative and goes to zero at about $x/D=10$, whereas the SBL streamwise pressure gradient is positive after $x/D=5$ and goes to zero around $x/D=13$.

\begin{figure*}
    \centering
    \includegraphics[width=0.95\textwidth]{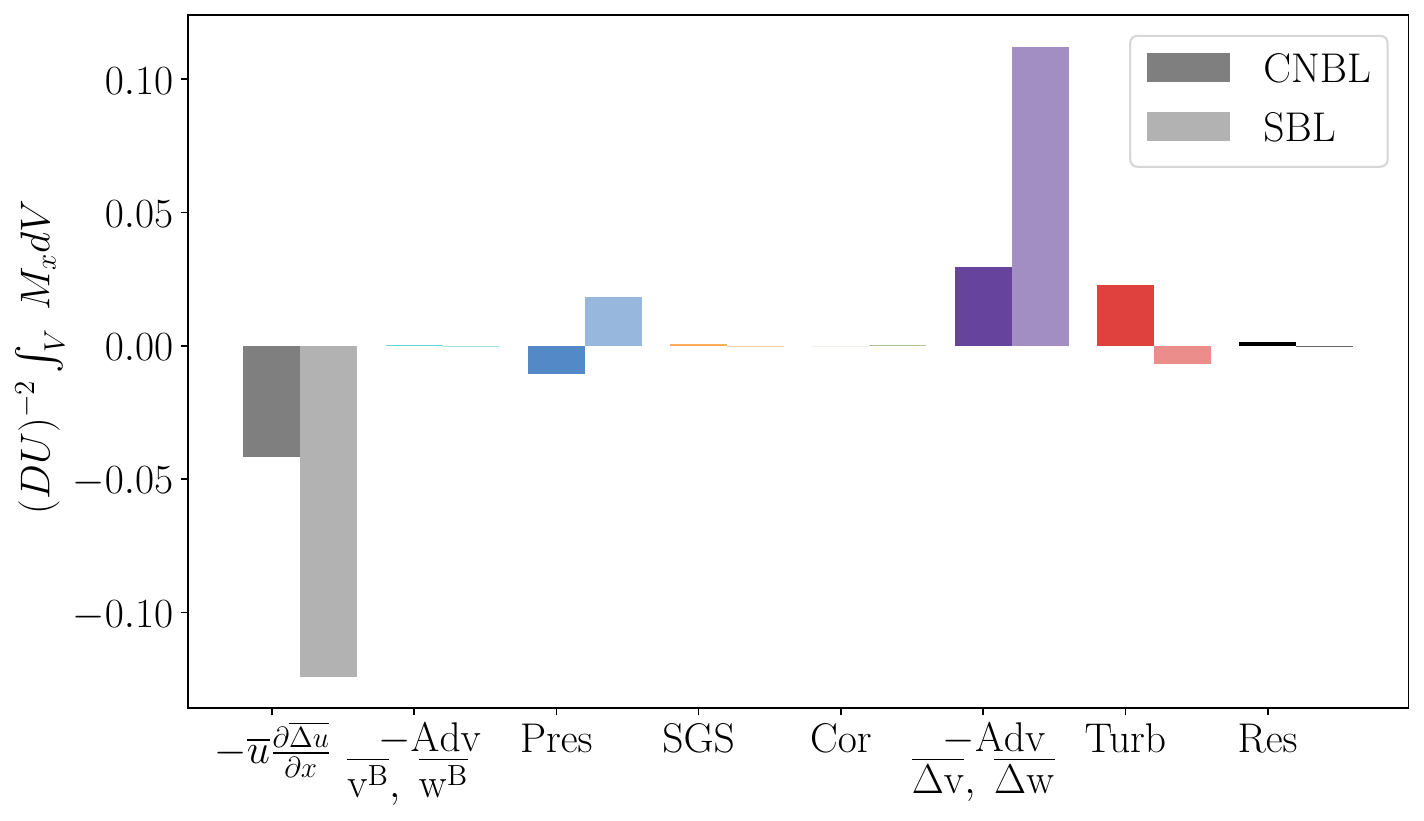}
    \caption{Streamwise momentum deficit budget integrated in a box control volume in the far wake ($x/D=5$ to $x/D=15$). Res denotes the budget residual.}\label{fig:nbl sbl xmom cv}
\end{figure*}

Again, we integrate the budgets---now in the box---to ascertain the overall effect of each term in the far wake. This is shown in Figure~\ref{fig:nbl sbl xmom cv}, where the importance of $\overline{\Delta v}$ and $\overline{\Delta w}$ advection is much more pronounced for both flows as compared with the streamtube-integrated budgets shown in Figure~\ref{fig:nbl sbl xmom streamtube}. This is particularly true for the SBL, which is now dominated by the deficit lateral and vertical advection in the balance with the streamwise advection. 

\subsection{\label{subsec:results tke} Wake TKE budget}
\subsubsection{\label{subsubsec:results tke a priori}\textit{A priori} wake analysis}

\begin{figure*}
    \centering
    \includegraphics[width=0.95\textwidth]{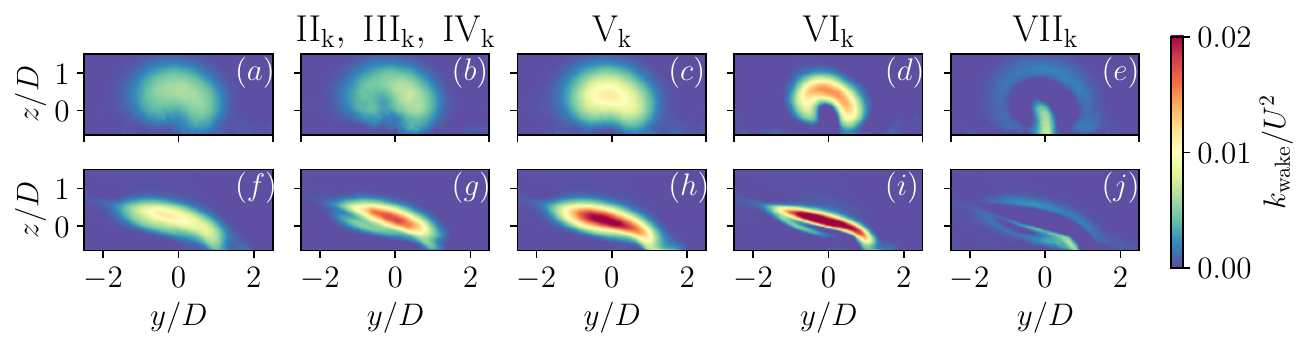}
    \caption{Slices of $k_\mathrm{wake}$ in the $y-z$ plane for the \textit{a priori} analysis at $x/D=10$. The CNBL is shown in the top row (Figs. ($a$)-($e$)) and the SBL is shown in the bottom row (Figs. ($f$)-($j$)). The title of each column indicates the terms that have been removed. No terms have been removed in Figs. ($a$) and ($f$); buoyancy, pressure and SGS transport have been removed in Figs. ($b$) and ($g$); dissipation has been removed in Figs. ($c$) and ($h$); turbulent transport has been removed in Figs. ($d$) and ($i$); and shear production has been removed in Figs. ($e$) and ($j$).}\label{fig:nbl sbl tke a priori}
\end{figure*}

Analogously to Section~\ref{subsubsec:results momentum a priori}, we integrate the wake-added TKE equation (Eq.~\ref{eqn:wake tke transport integral}) in the streamwise direction, and remove terms one by one. Terms II$_k$, III$_k$, and IV$_k$ are grouped together for the sake of clarity as these terms are all relatively small. The results of this \textit{a priori} analysis are shown in Fig.~\ref{fig:nbl sbl tke a priori} at a location of $x/D=10$. Figures~\ref{fig:nbl sbl tke a priori}($a$) and ($f$) show $k_\mathrm{wake}$ when no terms are removed. In comparing all other subfigures in Fig.~\ref{fig:nbl sbl tke a priori} to Figs.~\ref{fig:nbl sbl tke a priori}($a$) and ($f$), it is clear that all terms or groups of terms have an impact on the balance of wake-added TKE for both the CNBL and SBL. Removing dissipation (Fig.~\ref{fig:nbl sbl tke a priori}($c$) and ($h$)), turbulent transport (Fig.~\ref{fig:nbl sbl tke a priori}($d$) and ($i$)), and shear production (Fig.~\ref{fig:nbl sbl tke a priori}($e$) and ($j$)) has the largest impact on the predicted $k_\mathrm{wake}$. Without the dissipative mechanism, the maximum values of $k_\mathrm{wake}$ are too high, particularly for the SBL. Similarly, the magnitudes are too high when turbulent transport is removed, due to the lack of a crucially important diffusion mechanism in the flow. When shear production is removed, the magnitudes are too low as the main source of turbulence kinetic energy has been removed from the flow. 

\begin{figure*}
    \centering
    \subfloat[CNBL]{
    \includegraphics[width=0.45\textwidth]{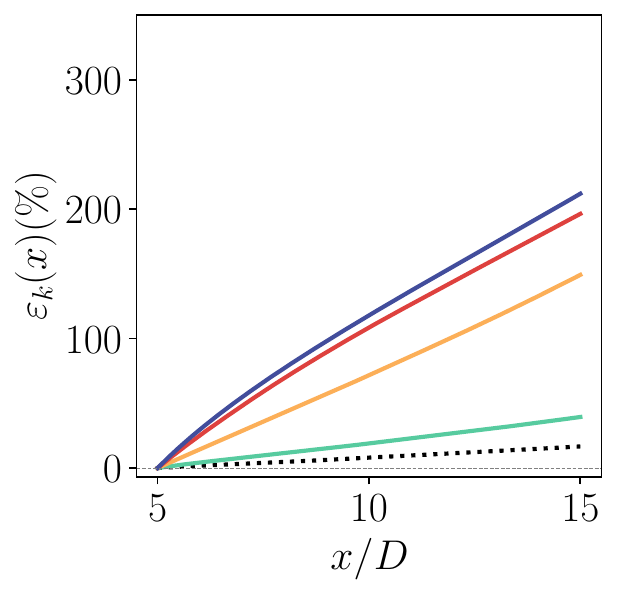}}
    \subfloat[SBL]{
    \includegraphics[width=0.45\textwidth]{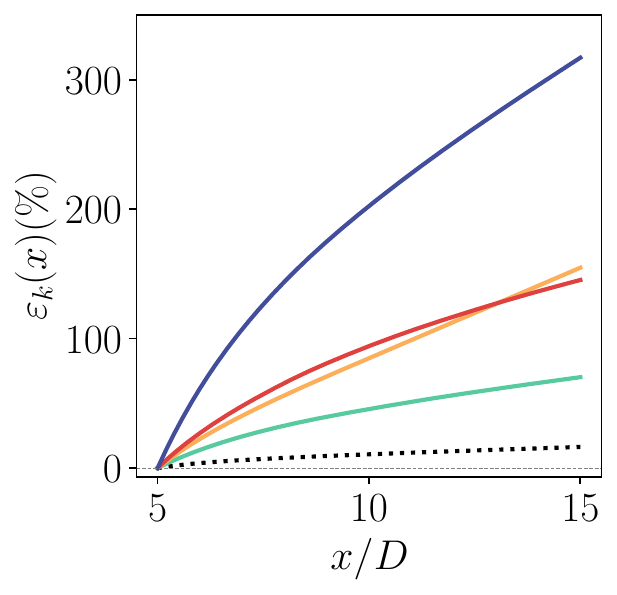}}
    \caption{Wake-added TKE error in $x$ as defined in Eq.~\ref{eq:wake tke error in x} for advection of $k_\mathrm{wake}$ by the deficit flow (II$_k$), buoyancy (III$_k$), and advection of $k^B$ by the deficit flow, pressure transport, SGS transport (IV$_k$)  removed (\errnodpdxsgscor[scale=0.5]); turbulent transport removed (\errnoRSdiv[scale=0.5]); shear production removed (\errnodeltavw[scale=0.5]); dissipation removed (\errnodiss[scale=0.5]);  and no terms removed (\errfullLES[scale=0.5]). The dashed gray line shows the location of zero error.}\label{fig:nbl sbl tke error in x}
\end{figure*}

As in Section~\ref{subsubsec:results momentum a priori}, we define an error metric given by
\begin{eqnarray}\label{eq:wake tke error in x}
    \varepsilon_k(x) = 100\% \times \frac{\lVert k_{\mathrm{wake},\;a\; priori}(x) - k_{\mathrm{wake},\;LES}(x)\rVert_2}{\lVert k_{\mathrm{wake},\;LES}(x)\rVert_2}.
\end{eqnarray}
The error in Eq.~\ref{eq:wake tke error in x} is shown in Fig.~\ref{fig:nbl sbl tke error in x}. As is evident in Fig.~\ref{fig:nbl sbl tke a priori}, removal of shear production produces the largest errors and removal of the combination of buoyancy, pressure and SGS transport produces the smallest error in both boundary layers. In the CNBL, turbulent transport displays a higher degree of importance through increased error over dissipation, while in the SBL these errors are much closer indicating a similar level of importance to the overall budget of $k_\mathrm{wake}$.
\subsubsection{\label{subsubsec:results tke budget analysis}Budget analysis}
We present budget analysis of wake-added TKE, similarly to Section~\ref{subsubsec:results momentum budget}, however, in place of control volume analysis, we instead look at $y-z$ slices of the budget terms in Eq.~\ref{eqn:wake tke transport} (as labeled in Eq.~\ref{eqn:wake tke transport integral}). This is done to simplify the analysis and present the spatial distribution of the wake-added TKE budget.

\begin{figure*}
    \centering
    \subfloat[$x/D=2.5$\label{fig:nbl sbl tke budget yz x/D=2.5}]{\includegraphics[width=0.95\textwidth]{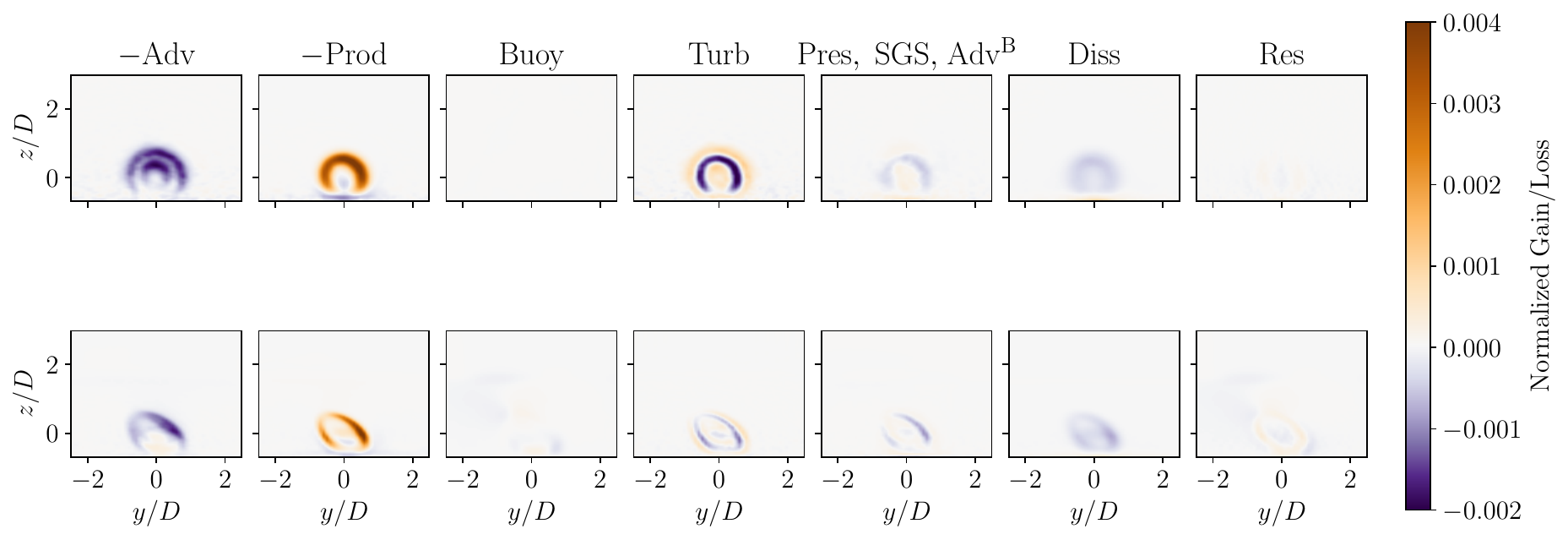}} \\
    \subfloat[$x/D=7.5$\label{fig:nbl sbl tke budget yz x/D=7.5}]{\includegraphics[width=0.95\textwidth]{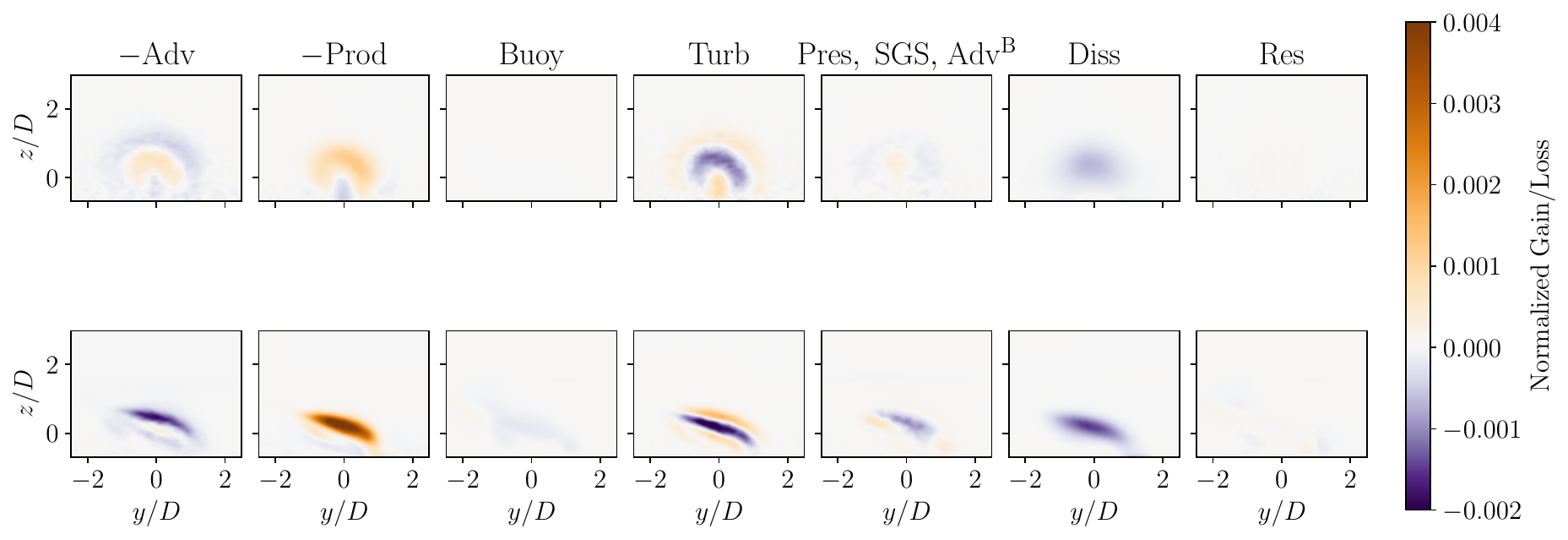}}
    \caption{Budget for $k_{\text{wake}}$ (Eq.~\ref{eqn:wake tke transport} in the $y-z$ plane at two streamwise locations. The top row in each subfigure is the budget for the CNBL and the bottom row is the budget for the SBL. All terms are normalized by $D/U^3$. Res refers to the residual, which is the sum of all the budget terms.}\label{fig:nbl sbl tke budget yz}
\end{figure*}
Figure~\ref{fig:nbl sbl tke budget yz} shows each $k_{\text{wake}}$ budget term from Eq.~\ref{eqn:wake tke transport} in the $y-z$ plane at $x/D=2.5$ (Fig.~\ref{fig:nbl sbl tke budget yz x/D=2.5}) and $x/D=7.5$ (Fig.~\ref{fig:nbl sbl tke budget yz x/D=7.5}). The first location roughly corresponds to the peak in production for the CNBL and the second location roughly corresponds to the peak in production for the SBL. For both flows, production is primarily balanced by advection, turbulent transport, and dissipation. Interestingly for the SBL, buoyant destruction is relatively negligible compared to dissipation. Buoyant destruction represents a direct effect of stratification on the wake, while the indirect effects are observed in the increased shear production and skewing of the wake in all terms. This is an important result as it illustrates that the main effects of stratification on the wake---in these stable conditions ($L=118$)---are from indirect changes to the ABL inflow. 

Of the four dominant terms identified above, three arise from nonlinear interactions in the velocity field, namely the production, advection, and turbulent transport. The base flow exactly satisfies the RANS equations, meaning that terms that derive from the nonlinear advection term in the Navier-Stokes equations are included in the transport equation for the wake deficit flow. In Section~\ref{subsec:results momentum}, advection of $\overline{\Delta u}$ by $\overline{\Delta v}$ and $\overline{\Delta w}$ is found to be important in the wake, particularly for the SBL. These terms involve interaction between the wake deficit and base flows, so here again we investigate the role of these terms and how it changes between the CNBL and SBL.

Starting with production---as given by term VII$_k$ in Eq~\ref{eqn:wake tke transport integral}---we can decompose the entire production term from Eq.~\ref{eqn:wake tke transport integral} into the various components involving different combinations of the base flow and the deficit flow. For clarity, we now denote the $k_\mathrm{wake}$ shear production term (term VII$_k$) as $\mathcal{P}_\mathrm{wake}$. Decomposing $\mathcal{P}_\mathrm{wake}$ into its various components we have
\begin{align}\label{eq:prod labels}
    \begin{split}
    \mathcal{P}_\mathrm{wake} = &-\overline{u_i'u_j'} \frac{\partial \overline{u_i}}{\partial x_j} + \overline{u_i\Bp u_j\Bp} \pdv{\overline{u_i^B}}{x_j} 
    \\=&\underbrace{-\;\overline{u_i\Bp u_j\Bp} \pdv{\overline{\Delta u_i}}{x_j}}_{\mathcal{P}_{BB\Delta}} \;\underbrace{-\;\overline{u_i\Bp \Delta u_j'} \pdv{\overline{u_i^B}}{x_j}}_{\mathcal{P}_{B\Delta B}} \;\underbrace{-\;\overline{\Delta u_i' u_j\Bp} \pdv{\overline{u_i^B}}{x_j}}_{\mathcal{P}_{\Delta B B}} \;\underbrace{-\;\overline{u_i\Bp \Delta u_j'} \pdv{\overline{\Delta u_i}}{x_j}}_{\mathcal{P}_{B\Delta \Delta}}
    \\&\underbrace{-\;\overline{\Delta u_i' u_j\Bp} \pdv{\overline{\Delta u_i}}{x_j}}_{\mathcal{P}_{\Delta B\Delta}}\;
    \underbrace{-\;\overline{\Delta u_i' \Delta u_j'} \pdv{\overline{u_i^B}}{x_j}}_{\mathcal{P}_{\Delta \Delta B}} \;\underbrace{-\;\overline{\Delta u_i' \Delta u_j'} \pdv{\overline{\Delta u_i}}{x_j} }_{\mathcal{P}_{\Delta \Delta \Delta}}.
    \end{split}
\end{align}
The labels under each term refer to the naming convention, where each $\Delta$ subscript corresponds to a wake deficit velocity and each $B$ subscript corresponds to a base flow velocity. Generally, this can be written as
\begin{equation}\label{eq:prod labels 2}
    \mathcal{P}_{\square \square \square} = -\overline{\square_i' \square_j'} \frac{\partial \overline{\square_i}}{\partial x_j}.
\end{equation}
We can similarly decompose the advection of $k_\mathrm{wake}$ (given by terms I$_k$, II$_k$, and the streamwise advection in Eq.~\ref{eqn:wake tke transport integral}) and turbulent transport of $k_\mathrm{wake}$ (given by term VI$_k$ in Eq.~\ref{eqn:wake tke transport integral}). For clarity, advection and turbulent transport of $k_\mathrm{wake}$ are denoted as $\mathcal{A}_\mathrm{wake}$ and $\mathcal{T}_\mathrm{wake}$, respectively. As with the production components in Eq.~\ref{eq:prod labels 2}, we label the components of advection as
\begin{equation}\label{eq:adv labels}
    \mathcal{A}_{\square \square \square} = -\frac{1}{2}\overline{\square_j} \pdv{\overline{\square_i' \square_i'} }{x_j}
\end{equation}
and the turbulent transport as
\begin{equation}\label{eq:turb labels}
    \mathcal{T}_{\square \square \square} = -\frac{1}{2}\pdv{}{x_j}\overline{ \square_j' \square_i' \square_i'}. 
\end{equation}
It is important to note that while 7 unique terms comprise $\mathcal{P}_\mathrm{wake}$, there are only 4 unique terms that comprise $\mathcal{A}_\mathrm{wake}$ and 5 unique terms that comprise $\mathcal{T}_\mathrm{wake}$. 

\begin{figure*}
    \centering
    \includegraphics[width=0.975\textwidth]{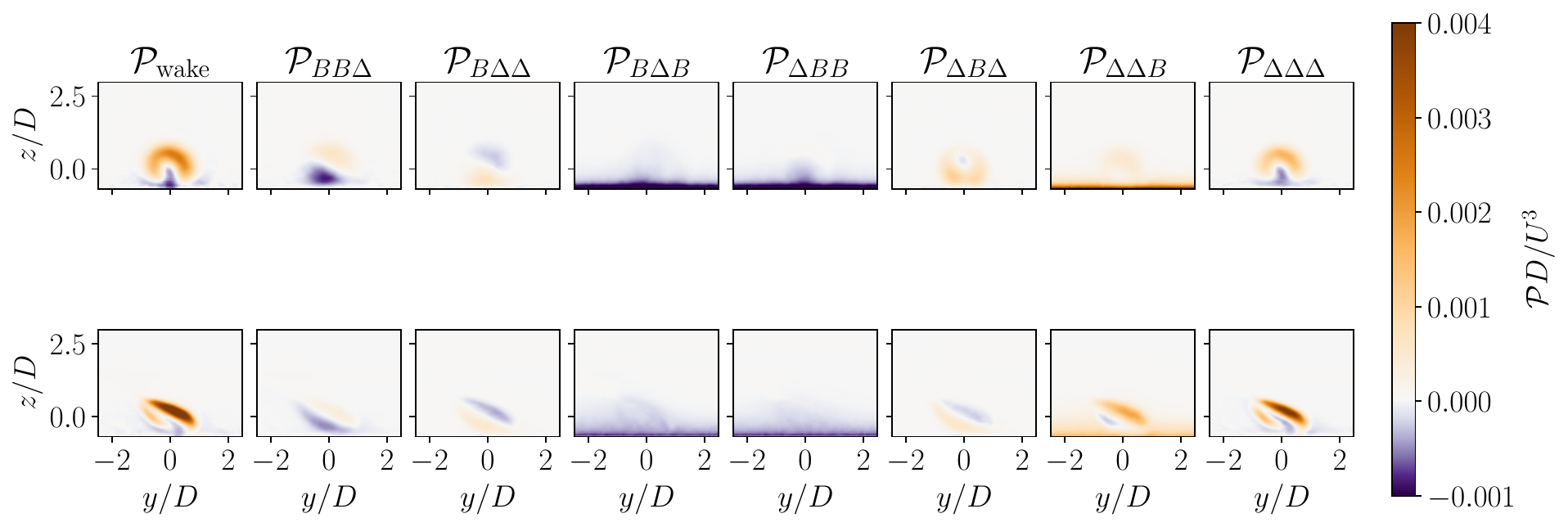}
    \caption{Components of the $k_{\text{wake}}$ production term as defined by Eq.~\ref{eq:prod labels} in the $y-z$ plane at $x/D=5$. $\mathcal{P}_{\text{wake}}$ refers to the entire wake TKE production term, which is the sum of all the terms to the right.}\label{fig:nbl sbl tke prod yz}
\end{figure*}

Figure~\ref{fig:nbl sbl tke prod yz} shows the 7 production terms that comprise $\mathcal{P}_{\text{wake}}$. For both flows, $\mathcal{P}_{\text{wake}}$ is primarily comprised of $\mathcal{P}_{\Delta \Delta B}$ and $\mathcal{P}_{\Delta \Delta \Delta}$. This first term is of particular interest because it is a combination of wake deficit Reynolds stresses $\overline{\Delta u_i' \Delta u_j'}$ and the gradient of the mean base velocity $\pdv{\overline{u_i^B}}{x_j}$. Clearly, this term is more pertinent for the SBL, given that in the CNBL it largely cancels with $\mathcal{P}_{B \Delta B}$ and $\mathcal{P}_{\Delta BB}$. Due to the increased velocity shear in the SBL base flow, the vertical gradient of the mean base flow is much larger than in the CNBL. As a result, terms that involve $\pdv{\overline{u^B}}{z}$ tend to have a larger effect on the wake in the SBL.

We can further quantify this result by computing the relative contribution of each component of production to the overall wake-added TKE production. We do this by taking the $\ell^2$ norm of the individual component $\mathcal{P}_{\square \square \square}$ and normalizing by the $\ell^2$ norm of $\mathcal{P}_\mathrm{wake}$ as shown in the equation below
\begin{eqnarray}\label{eq:norm}
    \overline{\eta}_{\mathcal{P}_{\square\square \square}} (x) = \left \langle \frac{\lVert\mathcal{P}_{\square \square \square}\rVert_2}{\lVert\mathcal{P}_\mathrm{wake} \rVert_2} \right \rangle_{box},
\end{eqnarray}
which is analogously defined for advection and turbulent transport of $k_\mathrm{wake}$. The brackets indicate that these terms are $y-z$ averaged in the box control volume. Because of the $\ell^2$ norm, the above metric does not necessarily sum to one when all the components are combined. As such, we define an additional metric in which Eq.~\ref{eq:norm} is normalized by the sum of each component. This is given by
\begin{eqnarray}\label{eq:norm2}
    \overline{\eta}^*_{\mathcal{P}_{\square\square \square}}(x) = \frac{\overline{\eta}_{\mathcal{P}_{\square\square \square}}(x)}{\sum_i \overline{\eta}_{\mathcal{P}_{i}}(x)},
\end{eqnarray}
where the denominator is the sum over all values of $\overline{\eta}_{\mathcal{P}_{\square \square \square}}$. Again, this metric is analogously defined for turbulent transport and mean advection.

\begin{figure*}
    \centering
    \subfloat[CNBL]{\includegraphics[height=0.35\textwidth]{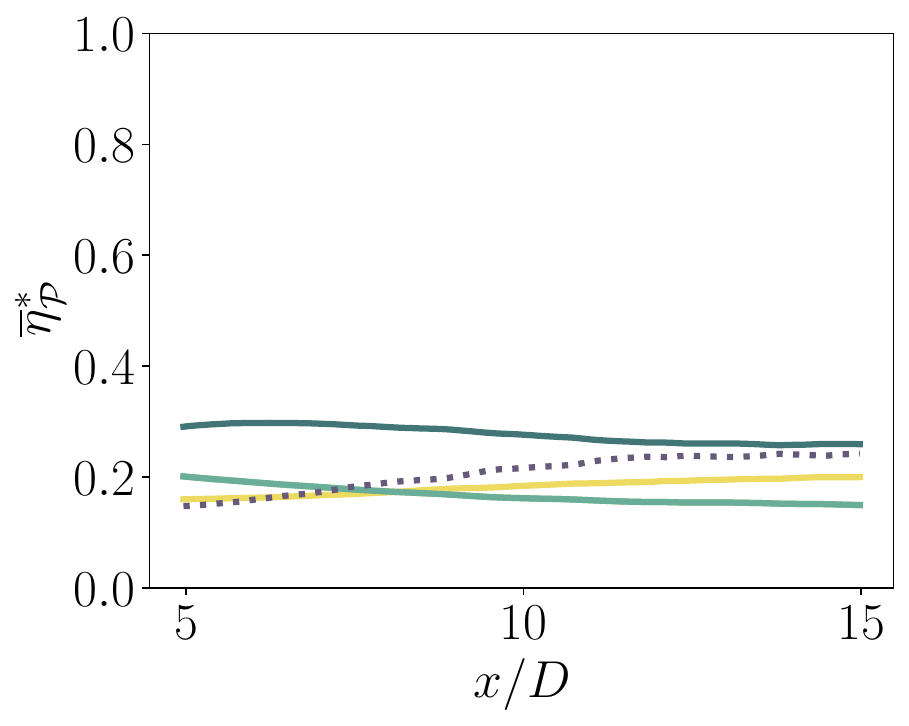}}
    \subfloat[SBL]{\includegraphics[height=0.35\textwidth]{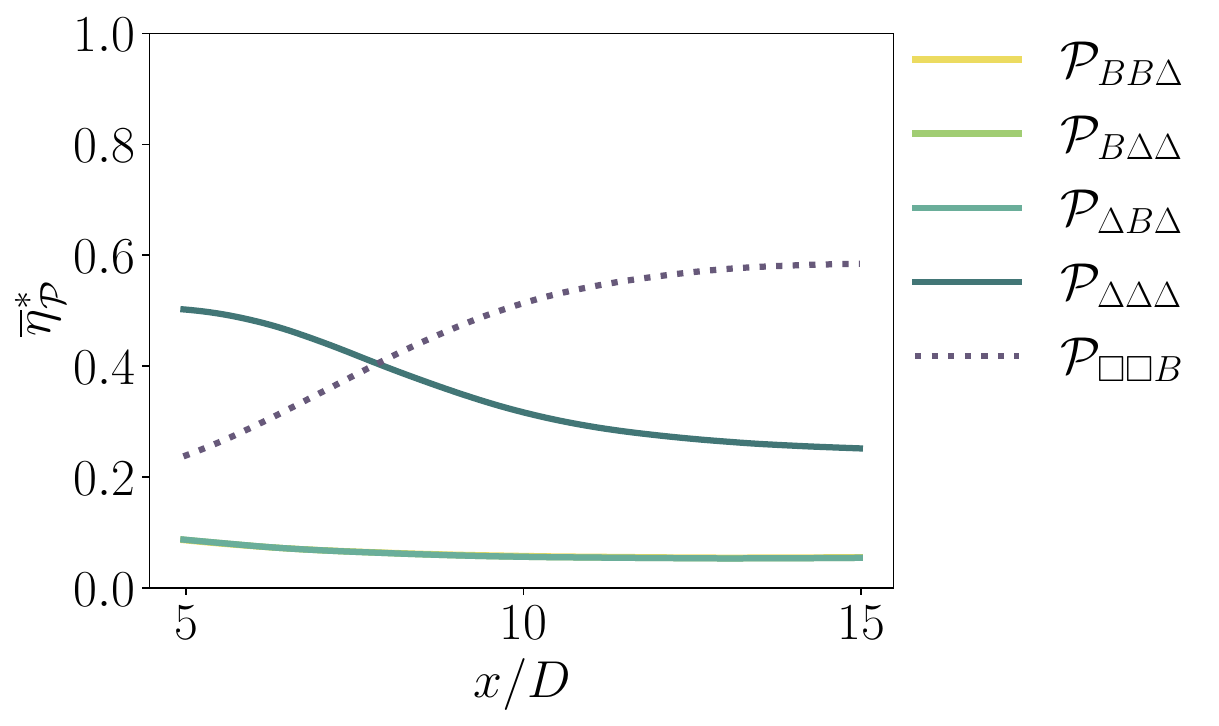}}
    \caption{Relative contribution of the components of production as defined in Eq.~\ref{eq:norm2}.}\label{fig:nbl sbl tke prod norm}
\end{figure*}

The metric in Eq.~\ref{eq:norm2} is shown for production in Fig.~\ref{fig:nbl sbl tke prod norm}. All terms that contain a base flow gradient are combined into one term denoted by $\mathcal{P}_{\square \square B}$. Figure~\ref{fig:nbl sbl tke prod norm} clearly shows that $\mathcal{P}_{\square \square B}$ is more important to the overall $\mathcal{P}_{\mathrm{wake}}$ in the SBL than in the CNBL, where after about 8 diameters downstream of the turbine, this term becomes dominant. We also observe that the terms that contain base flow fluctuations are larger in the CNBL than in the SBL.  

\begin{figure*}
    \centering
    \includegraphics[width=0.95\textwidth]{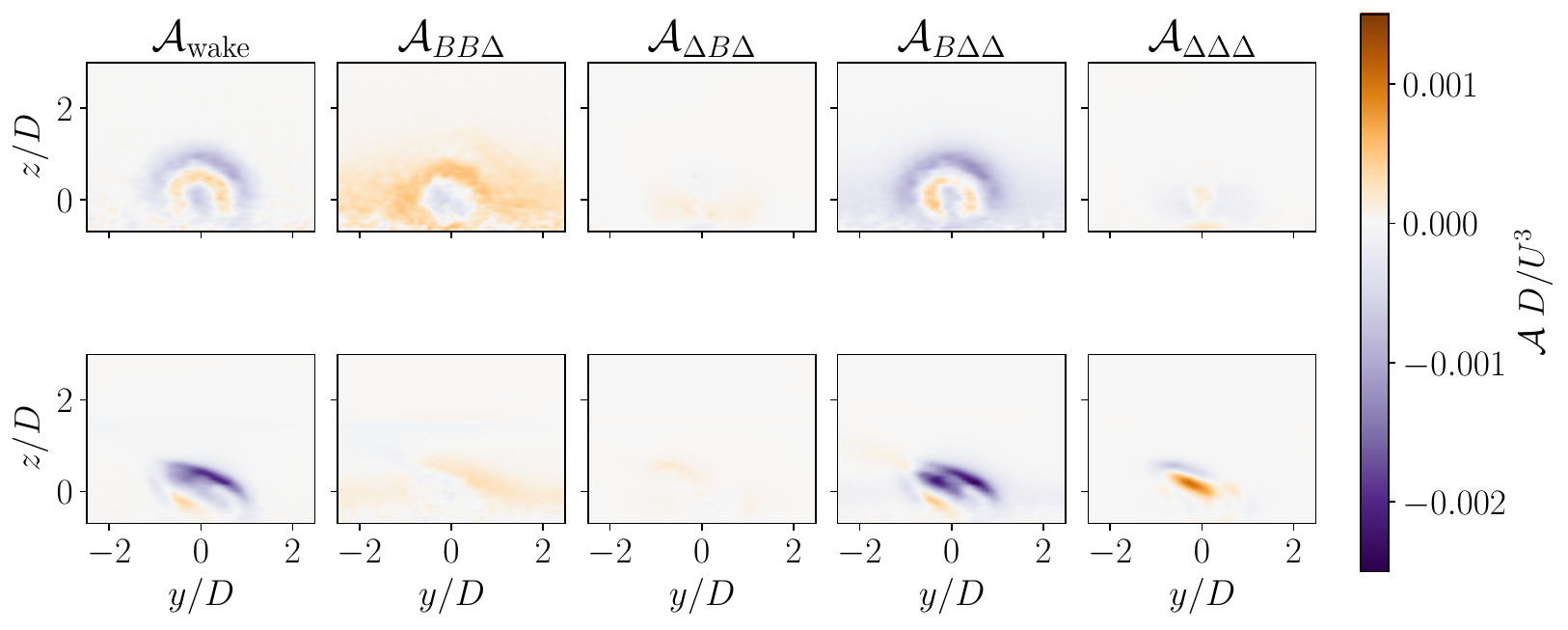}
    \caption{Components of the $k_{\text{wake}}$ production term as defined by Eq.~\ref{eq:adv labels} in the $y-z$ plane at $x/D=5$. $\mathcal{A}_{\text{wake}}$ refers to the entire wake TKE production term, which is the sum of all the terms to the right.}\label{fig:nbl sbl tke adv yz}
\end{figure*}

Similar to the production, the advection and turbulent transport terms in the $k_{\text{wake}}$ budget involve correlations between base flow and wake deficit flow fluctuations. Figure~\ref{fig:nbl sbl tke adv yz} shows the components of the advection of wake-added TKE $\mathcal{A}_{\text{wake}}$ as given by Eq.~\ref{eq:adv labels}. In both flows, $\mathcal{A}_{B\Delta \Delta}$ is the dominant term. For the CNBL, there is also a substantial opposing contribution from $\mathcal{A}_{BB\Delta}$, which is advection of the correlation between $u_i\Bp$ and $\Delta u_i'$ by the base flow $\overline{u^B}$. In comparison, this term is relatively negligible in the SBL. This is likely due to the fact that the turbulence content---based on either TKE or turbulence intensity---is higher in the CNBL. 

\begin{figure*}
    \centering
    \subfloat[CNBL]{\includegraphics[height=0.35\textwidth]{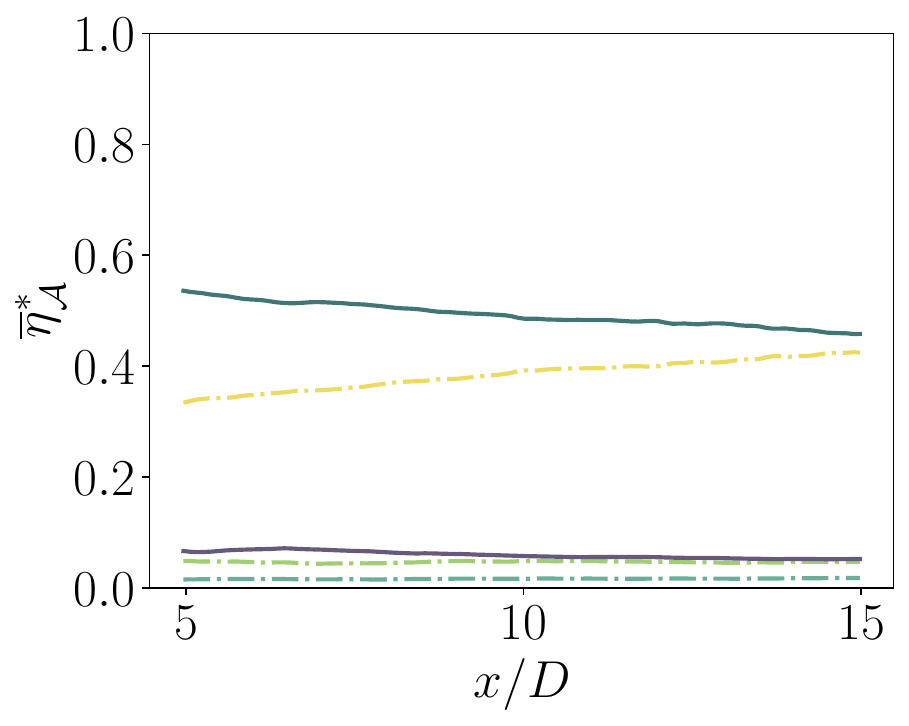}}
    \subfloat[SBL]{\includegraphics[height=0.35\textwidth]{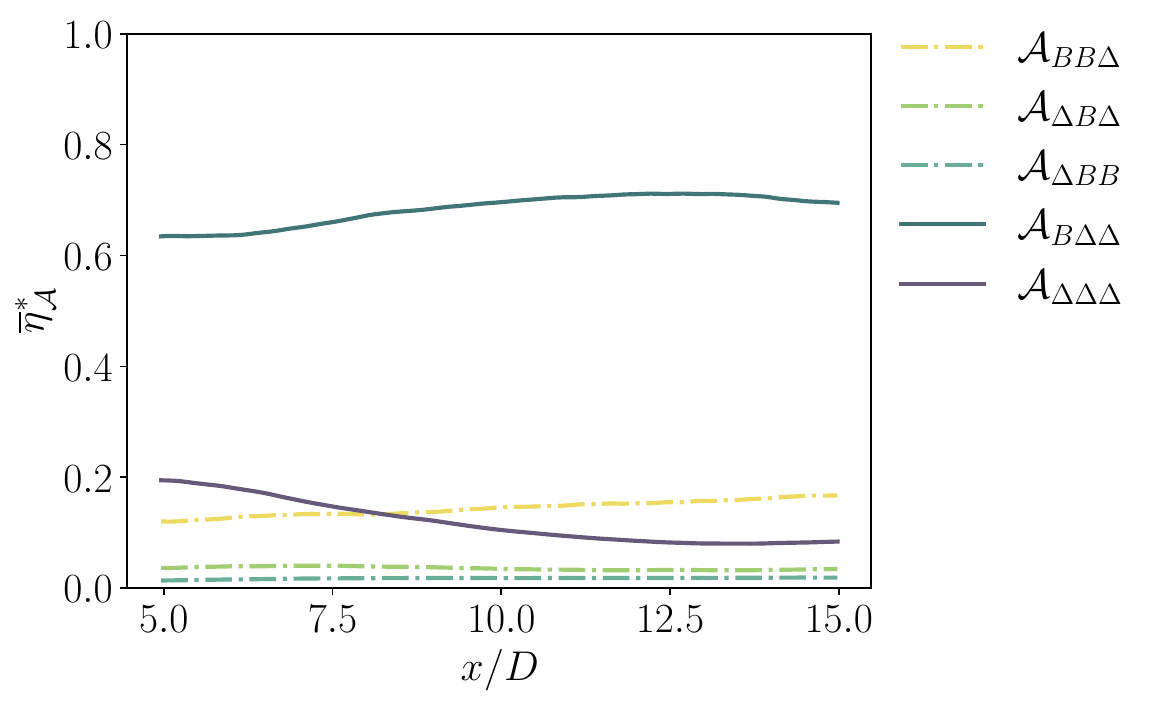}}
    \caption{Relative contribution of the components of advection as defined in Eq.~\ref{eq:norm2}.}
    \label{fig:nbl sbl tke adv norm}
\end{figure*}

Again, we look at the metric defined in Eq.~\ref{eq:norm}, now for advection, in Fig.~\ref{fig:nbl sbl tke adv norm}. All terms with base flow fluctuations have been grouped for clarity into the term denoted $\mathcal{A}_{\square \square B}$. This term is clearly much more important to the overall advection in the CNBL case. 


\begin{figure*}
    \centering
    \includegraphics[width=0.95\textwidth]{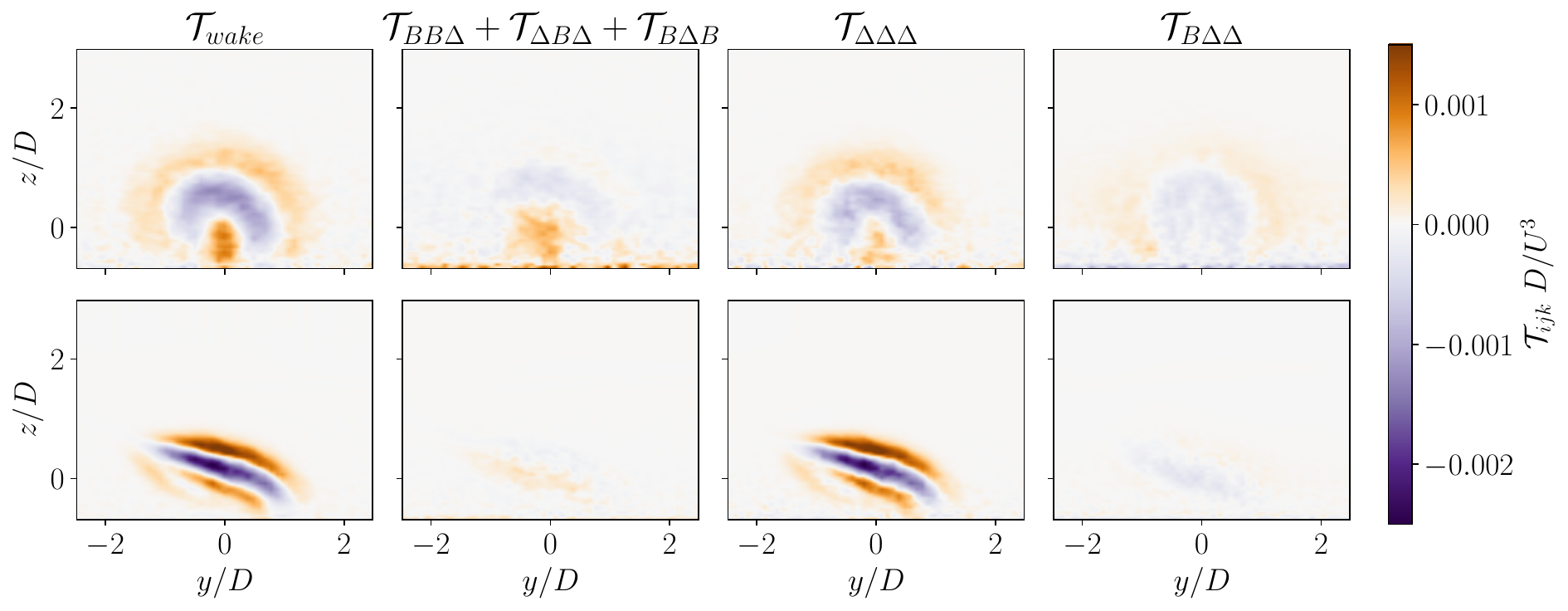}
    \caption{Components of the $k_{\text{wake}}$ turbulent transport term as defined by Eq.~\ref{eq:turb labels} in the $y-z$ plane at $x/D=7.5$. $\mathcal{T}_{\text{wake}}$ refers to the entire wake TKE turbulent transport term, which is the sum of all the terms to the right.}\label{fig:nbl sbl tke turb trans yz}
\end{figure*}

\begin{figure*}
    \centering
    \subfloat[CNBL]{\includegraphics[height=0.3\textwidth]{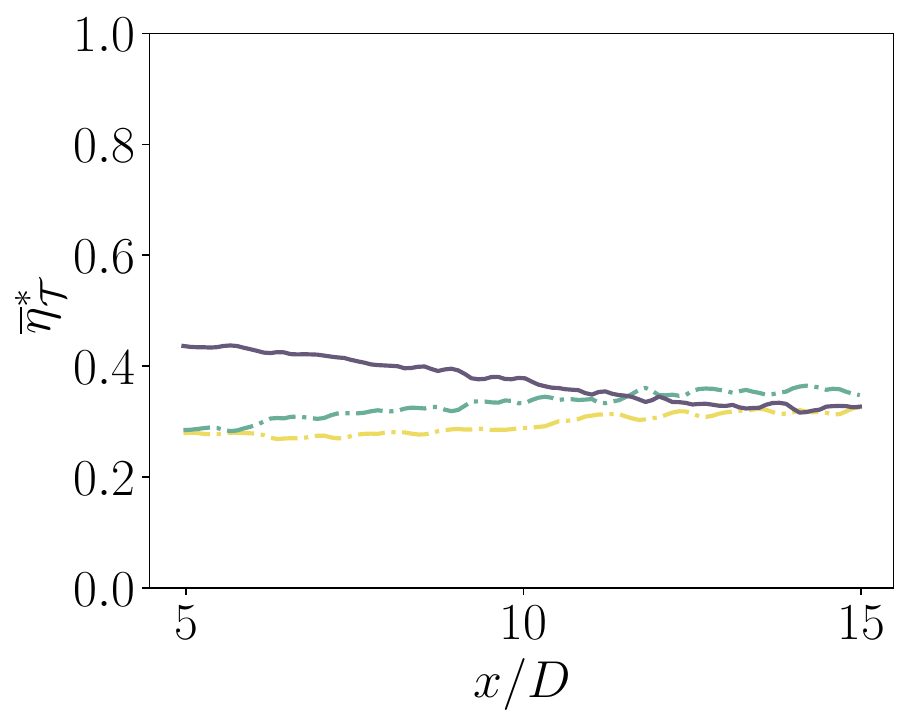}}
    \subfloat[SBL]{\includegraphics[height=0.3\textwidth]{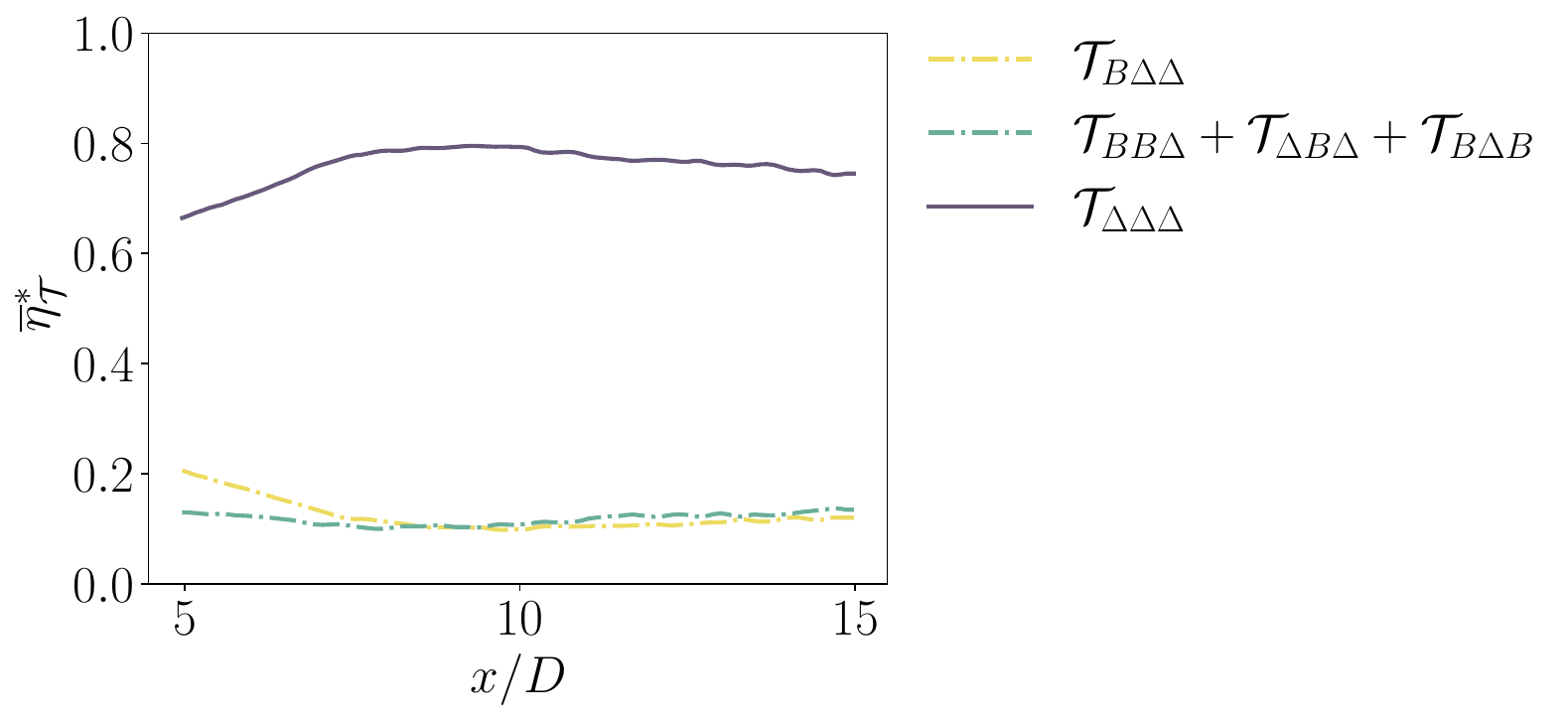}}
    \caption{Relative contribution of the components of turbulent transport as defined in Eq.~\ref{eq:norm2}.}\label{fig:nbl sbl tke turb norm}
\end{figure*}

The role of the base flow turbulence can be further seen in the turbulent transport of $k_{\text{wake}}$ in Figs.~\ref{fig:nbl sbl tke turb trans yz} and \ref{fig:nbl sbl tke turb norm}. In the SBL, $\mathcal{T}_{\text{wake}}$ predominantly comes from $\mathcal{T}_{B \Delta \Delta}$, which involves only the correlation between wake deficit velocity fluctuations. While this is also true for the CNBL, the neutral case also exhibits dependence on the other four terms, which all involve the correlation of wake deficit and base flow fluctuations. Again, this indicates that $k_{\text{wake}}$ in the CNBL interacts with the base flow primarily through correlation with the base flow fluctuations, which is in contrast to the SBL, which interacts with the base flow primarily through the mean base flow gradients.

The analysis of the terms that contain interactions between the base flow and the deficit flow in the $k_{\mathrm{wake}}$ budget illustrates how the indirect effects of stratification impact wake-added TKE. As discussed above, the direct effect of stratification felt through the buoyant destruction term has a relatively small contribution to the $k_\mathrm{wake}$ balance in comparison to other destructive mechanisms, such as dissipation. As such, what has been identified in this section provides insight into the primary effects of stratification on wake turbulence: the contributions of the base flow to the wake-added TKE budget is felt through the base flow fluctuations, which are linked to base flow turbulence intensity, and the base flow gradients, which are linked to base flow direction and speed shear. The former is dominant in the neutral case, while the latter is dominant in the stable case. In the stable case, these indirect effects dominate the direct buoyant destruction of TKE for these stable ABL conditions.

\subsection{\label{subsec:discussion} Evaluation of wake-added TKE modeling}

In the preceding sections, we analyzed the streamwise momentum and turbulence kinetic energy budgets for the wake deficit flow to better understand the physical mechanisms that govern wake dynamics. We focus on wake quantities to better parse the direct effects of base flow atmospheric stratification from the indirect effects of atmospheric stratification. This exploration is pertinent from a flow physics perspective and offers insight into modeling considerations as well. As previously stated, wake models approximate the mean streamwise wake deficit velocity $\overline{\Delta u}$, often through linearized, analytical models. These models are typically developed for uniform~\cite{bastankhah_new_2014} or neutrally stratified inflow~\cite{niayifar_analytical_2016}. Wake turbulence in these models is incorporated through a wake spreading parameter, which is often treated as a linear function of the total streamwise turbulence intensity $I_T$, which is a combination of the wake-added turbulence intensity $I_+$ and the ambient turbulence intensity $I_0$~\cite{niayifar_analytical_2016,stevens_flow_2017}. Wake-added turbulence intensity is defined as
\begin{equation}\label{eq:I+}
    I_+ = \sqrt{I_T^2 - I_0^2},
\end{equation}
where $I_T$ is the total turbulence intensity in the wake $I_0$ is the ambient streamwise turbulence intensity. The total horizontal turbulence intensity in the wake is defined as
\begin{equation}\label{eq:Iwake}
    I_T = \sqrt{\overline{u'u'} + \overline{v'v'}}/u_h,
\end{equation}
where $u_h$ is the hub height velocity. Currently in most analytical wake models, the horizontal or streamwise turbulence intensity is used in a model setting and is related to total turbulence intensity through the removal of the vertical fluctuations. In many cases, the vertical fluctuations are found to be negligible and so the total streamwise turbulence intensity and total turbulence intensity are approximately equal~\cite{klemmer_evaluation_2024}. The ambient turbulence intensity is similarly defined for the base flow fluctuations. In many cases, $I_+$ is empirically modeled via the Crespo-Hern\'andez model~\cite{crespo_turbulence_1996} given by
\begin{equation}\label{eq:crespo}
    I_+ = 0.73a^{0.8325}I_0^{-0.0325} (x/D)^{-0.32},
\end{equation}
where a is the induction factor. The induction factor has been calculated for both flows as
\begin{equation}\label{eq:induction}
    a = 1 - \frac{u_d}{u^B},
\end{equation}
where $u_d$ is the rotor-averaged velocity at the disk and $u^B$ is the rotor-averaged base flow velocity. The ambient turbulence intensity input to the model is calculated a hub height. Note that the negative exponent on $I_0$ in Eq.~\ref{eq:crespo} is a departure from what is commonly used in literature, but it has been identified as the intended model constant by Zehtabiyan-Resaie and Abkar~\cite{zehtabiyan-rezaie_short_2023}.

\begin{figure*}
    \centering
    \includegraphics[width=0.8\textwidth]{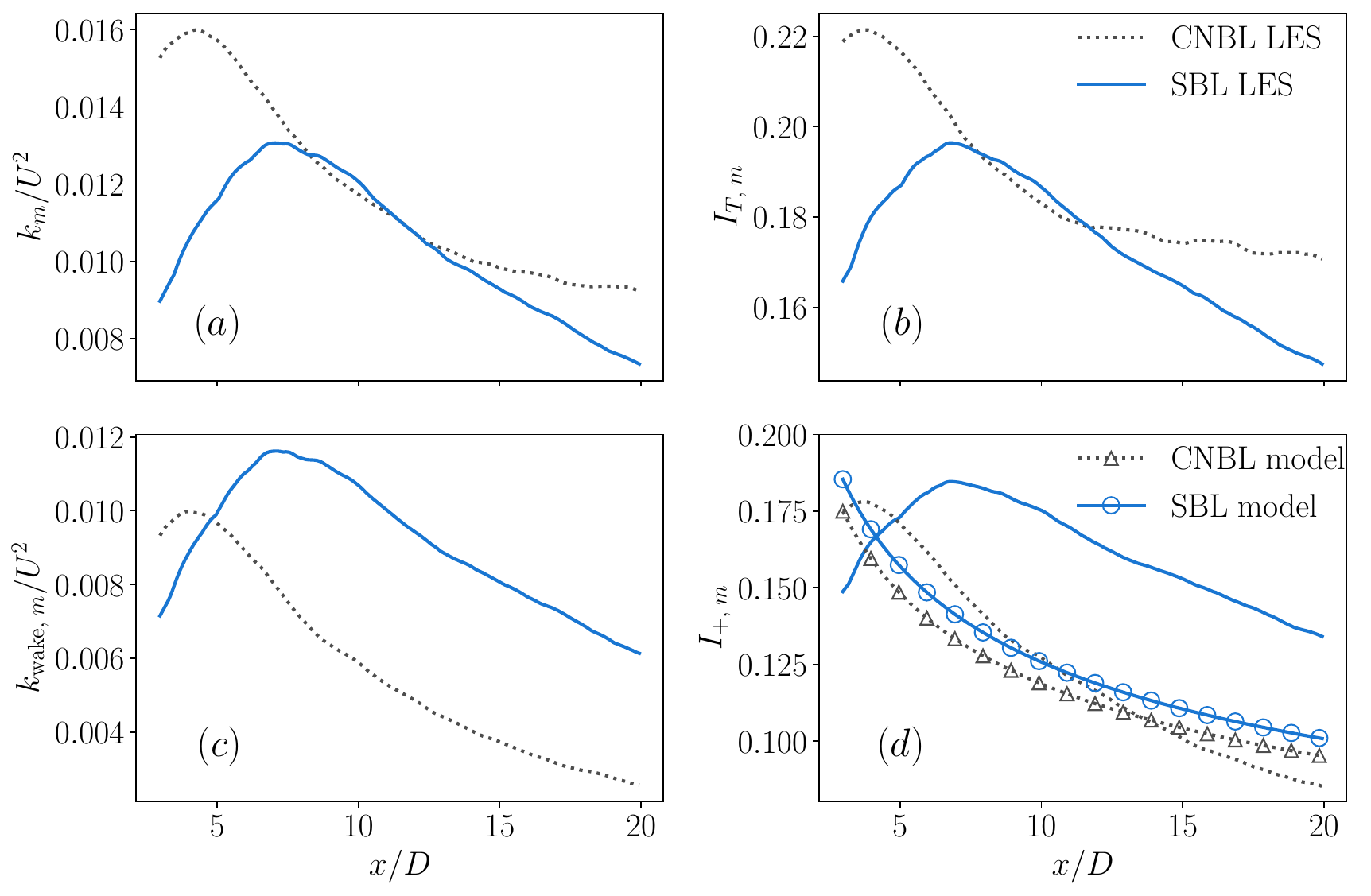}
    \caption{Maximum values of $k$ $(a)$, $I_T$ $(b)$, $k_{\text{wake}}$ $(c)$, and $I_+$ $(d)$ at each $x$ location in the wake. The maximum value of $I_+$ is compared with the model for $I_+$ from the Crespo-Hern\'andez turbulence model in Eq.~\ref{eq:crespo}.}\label{fig:nbl sbl wake added turbulence}
\end{figure*}

In Section~\ref{subsec:results tke}, we have analyzed $k_\text{wake}$, which similarly provides information about the turbulent energy content in the wake. These two quantities are related through the following expression
\begin{equation}\label{eq:I+ kwake}
    I_+ = \sqrt{2k_\text{wake} - \overline{w'w'}_\text{wake}}/u_h,
\end{equation}
where $\overline{w'w'}_\text{wake} = \overline{w'w'} - \overline{w\Bp w\Bp}$. Figure~\ref{fig:nbl sbl wake added turbulence} shows the maximum values of $k_\text{wake}$ and $I_+$ at each streamwise location in the wake, as well as the maximum values of $k$ and $I_T$. We take the maximum values as the model in Eq.~\ref{eq:crespo} was originally tuned to match the maximum $I_+$~\cite{crespo_turbulence_1996}. Looking first at Fig.~\ref{fig:nbl sbl wake added turbulence}(c), we find that the CNBL and SBL differ in the location of the peak of the maximum value of $k_\text{wake}$. This result is expected given that TKE is found to peak further downstream for stable flows as compared with unstable and neutral flows~\cite{wu_new_2023}. This due in part to the higher $I_0$ in neutral flows. Comparing this with $k_m$ in Fig.~\ref{fig:nbl sbl wake added turbulence}, we find that the peak locations are consistent between $k$ and $k_\text{wake}$.

Figure~\ref{fig:nbl sbl wake added turbulence}(d) shows a comparison of the maximum $I_{+}$ from the LES with $I_+$ from the model given by Eq.~\ref{eq:crespo} for both flows. The wake-added turbulence intensity exhibits the same behavior as $k_\text{wake}$ in Fig.~\ref{fig:nbl sbl wake added turbulence} (c), where the peak in the SBL is further downstream than that of the CNBL. Additionally, the magnitude of both $k_\text{wake}$ and $I_+$ is higher in the SBL after the peak values in the CNBL at around $x/D=5$. This is likely due to two factors. The first is the increased $I_0$ in the CNBL. Overall, the CNBL has higher ambient turbulence intensity at hub height with $I_0=0.13$, while $I_0=0.07$ in the SBL. Previous studies have found that lower ambient turbulence intensity leads to higher wake-added turbulence intensity in the far wake~\cite{chamorro_windtunnel_2009,wu_atmospheric_2012a,wu_effects_2020}. In an experimental study of marine turbines with different ambient turbulence levels~\cite{mycek_experimental_2014}, Mycek et al. found that in wakes with lower $I_0$ the shear layer that forms grows unperturbed by the ambient turbulence. When $I_0$ is higher, the distance at which the shear layer persists is lower due to the enhanced mixing from higher ambient turbulence. The prolonged shear layer in the lower $I_0$ case allows for increased production of turbulence behind the rotor, thus resulting in increased turbine-induced or wake-added turbulence intensity in this case. 

The second factor that likely gives rise to higher $I_+$ in the SBL is the presence of a LLJ. LLJs have been shown to increase entrainment of turbulence kinetic energy in the wake of stable boundary layers when the LLJ is above the wind farm~\cite{gadde_interaction_2021}, as is the case in the present study (see Fig.~\ref{fig:ws comp}). In the wake region, the SBL wake entrains turbulent energy from the LLJ. This coupled with the increased shear in the SBL relative to the CNBL, can lead to higher wake-added turbulence.

Despite the differences observed in Fig.~\ref{fig:nbl sbl wake added turbulence}, the model predicts very similar values of $I_+$ in both cases. The model in Eq.~\ref{eq:crespo} is highly dependent on the induction factor, which only differs by 3\% between the CNBL and SBL. Even though $I_0$ differs by about 60\%, the model for $I_+$ depends very little on the ambient turbulence intensity. While the overall trend is captured by the model, particularly for the CNBL, the model has no knowledge of stability and is unable to capture the difference in the magnitude and location of the peaks for the two flows or the increased magnitude of $I_+$ in the SBL.  

\begin{figure*}
    \centering
    \subfloat[$x-y$ plane at hub height\label{fig:I_wake xy}]{\includegraphics[width=0.45\textwidth]{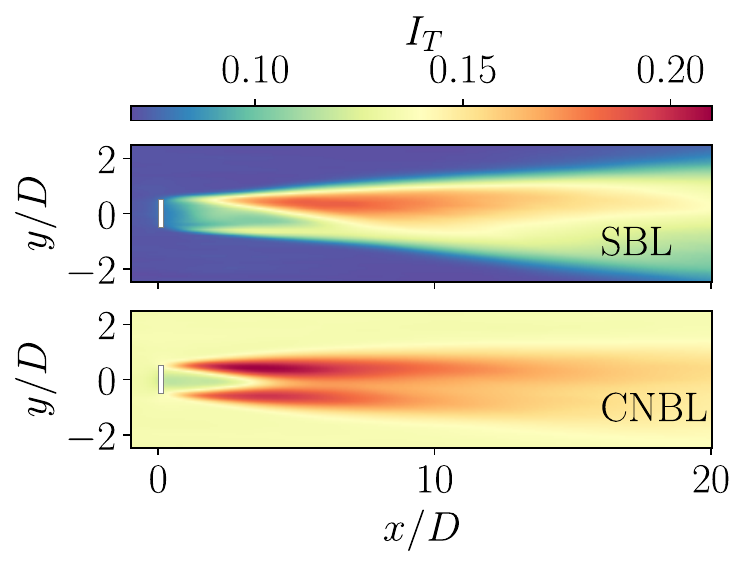}}
    \subfloat[$x-z$ plane centered on the turbine\label{fig:I_wake xz}]{\includegraphics[width=0.45\textwidth]{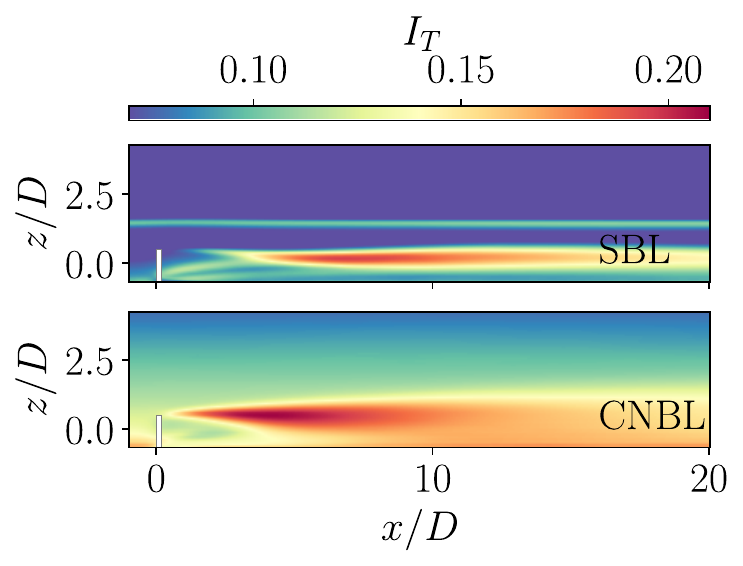}}
    \caption{Comparison of the $I_T$ at hub height in the $x-y$ plane (a) and centered on the turbine in the $x-z$ plane (b).}\label{fig:I_wake comp}
\end{figure*}
Ultimately, this analysis reveals two important things. First, state-of-practice turbulence models are unable to capture differences in stratified flows due to a lack of knowledge of both the direct and indirect effects of atmospheric stratification on wake physics. This may not be limited to stratification but also could apply to any flows that deviate from the uniform or neutral profiles that these models were designed for. Second, differences in the flow physics that arise from stratification significantly affect both quantities of interest, such as $\overline{\Delta u}$, and model quantities like $I_+$. In the SBL, both the increased shear in the base flow and the interaction of the wake with LLJ give rise to increased wake-added turbulence relative to the CNBL. These findings provide insight into these effects of stratification, which can be used to address some of the present shortcomings in both the model form and tuned parameters of state-of-practice engineering wake models when used in non-neutral stratified flows. Coupling the present study with an analysis of the models can serve to aid in the development of models that are robust to realistic atmospheric conditions. 

\section{\label{sec:conclusions} Conclusions}
We have presented analysis of turbine wake flow physics in two atmospheric boundary layers: a conventionally neutral boundary layer and a stable boundary layer. Specifically, we have analyzed the wake-decomposed flow fields by subtracting the base flow from the full flow field to isolate the wake and then performed turbulent budget analysis on the streamwise wake deficit momentum and the wake-added turbulence kinetic energy. For the streamwise wake deficit momentum, we performed an \textit{a priori} analysis in which the wake deficit was computed via integration of the governing transport equation using LES data. To elucidate the quantitative effect of each physical mechanism, terms were removed one at a time to observe the effect this had on the wake. To complement this analysis, we also computed the streamtube and box control volume integrated budgets. For the wake-added turbulence kinetic energy, we also performed the \textit{a priori} analysis and additionally analyzed the terms that arise from the interaction between the base flow and the wake deficit flow to better understand how the indirect effects of stratification felt through the base flow influence the wake deficit. Through this budget analysis, we have identified the important physical mechanisms in the wake of a single turbine under neutral and stable stratification. 

Throughout this work, we have explored the role of both the direct and indirect effects of stratification on wind turbine wakes. For both the neutral and the stable boundary layers, the indirect effects are dominant. These effects are felt through the base flow, which interacts with the wake deficit flow through the nonlinear advection term in the Navier-Stokes equations. For the neutral case, the primary indirect effect is the increased turbulence intensity in the base flow relative to the stable base flow. The dominant indirect effect on the stable flow, is the increased degree of both speed and direction shear, which is found to have a much more significant impact on both the streamwise momentum deficit and the wake-added turbulence kinetic energy than the direct buoyancy forcing. Together, these indirect effects of stratification act to alter the physical mechanisms that dominant the streamwise momentum balance and energy balance in the wake. The implication of this is seen in the evaluation of the model for the wake-added turbulence intensity. In comparing the wake-added turbulence intensity from the LES data with the wake-added turbulence intensity calculated via the analytical Crespo-Hern\'andez model, we found that while the model displayed reasonable agreement in the neutral case, in the stable case the location and magnitude of the peak are both incorrectly predicted. These results illustrate that the current state-of-practice models that lack knowledge of stratification do not adequately capture the indirect effects of atmospheric stability.

 Overall, these results indicate that the indirect effects of stratification greatly affect the structure of the wake of a single turbine. These differences manifest primarily in the interaction between the wake and the base flow, making it difficult to entirely isolate the wake deficit from the incident base flow. With this in mind, it is important to revisit the models that are currently used to model wind turbine wakes given that they typically are stratification-agnostic or employ \textit{ad hoc} modifications that lack wide applicability. As in all modeling, it is up to the modeler to decide their tolerance for uncertainty. However, with the trend of increasing rotor diameter, model assumptions that were previously adequate need to be re-examined. As such, the present analysis provides an initial exploration of the governing physics in the wake both to add to the growing body of knowledge on stratified atmospheric boundary layer flows and to help inform future modeling endeavors. 

\section{Acknowledgements}
The authors gratefully acknowledge support from the National Science Foundation (Fluid Dynamics program, grant number FD-2226053, Program Manager: Dr. Ronald D. Joslin) and Siemens Gamesa Renewable Energy. All simulations were performed on Stampede2 and Stampede3 supercomputers under the NSF ACCESS project ATM170028.

\bibliography{main}

\end{document}